\documentclass[twocolumn,floatfix,prb,aps,showpacs]{revtex4}

\usepackage{graphicx,color,amsmath,amssymb}
\usepackage{nicefrac}

\newcommand{\PRL}{Phys.\ Rev.\ Lett.}
\newcommand{\PRB}{Phys.\ Rev.\ B}

\newcommand{\NPB}{Nucl.\ Phys.\ B}
\newcommand{\Ann}{Ann.\ of\ Phys.}
\newcommand{\JPA}{J. Phys. A}
\newcommand{\be}{\begin{equation}}
\newcommand{\ee}{\end{equation}}

\begin{document}

\title{Tripartite Composite Fermion States}
\author{G. J. Sreejith,$^{1,2}$ Ying-Hai Wu,$^2$ A. W\'ojs,$^3$ and J. K. Jain$^2$}
\affiliation{$^1$NORDITA, Roslagstullsbacken 23, 10691 Stockholm, Sweden}
\affiliation{$^2$Department of Physics, 104 Davey Lab, Pennsylvania State University, University Park PA, 16802, USA}
\affiliation{$^3$Institute of Physics, Wroclaw University of Technology, 50-370 Wroclaw, Poland}

\date{\today}
\begin{abstract}
The Read-Rezayi wave function is one of the candidates for the fractional quantum Hall effect at filling fraction $\nu=2+\nicefrac{3}{5}$, and thereby also its hole conjugate at $2+\nicefrac{2}{5}$. We study a general class of ``tripartite" composite fermion wave functions, which reduce to the Read-Rezayi ground state and quasiholes for appropriate quantum numbers, but also allow a construction of wave functions for quasiparticles and neutral excitations by analogy to the standard composite fermion theory. We present numerical evidence in finite systems that these trial wave functions capture well the low energy physics of a 4-body model interaction. We also compare the tripartite composite fermion wave functions with the exact Coulomb eigenstates at $2+\nicefrac{3}{5}$, and find reasonably good agreement. The ground state as well as several excited states of the 4-body interaction are seen to evolve adiabatically into the corresponding Coulomb states for $N=15$ particles. These results support the plausibility of the Read-Rezayi proposal for the $2+\nicefrac{2}{5}$ and $2+\nicefrac{3}{5}$ fractional quantum Hall effect. However, certain other proposals also remain viable, and further study of excitations and edge states will be necessary for a decisive establishment of the physical mechanism of these fractional quantum Hall states. 
\pacs{73.43.-f, 05.30.Pr, 71.10.Pm}
\end{abstract}
\maketitle

\section{Introduction}

The richness of the fractional quantum Hall effect (FQHE) is reflected by the fact that there exist evidences for $\sim 75$ fractions so far \cite{WeiPan}.  In spite of tremendous progress during the last three decades, the physical origin of certain FQHE states is still under debate. A prominent example is the origin of the FQHE at $\nicefrac{12}{5}$ or $\nicefrac{13}{5}$ in GaAs \cite{Kang,PanXia2008,PanXiafivehalf1999,PanXia2004} where, in spite of several proposals \cite{ReadRezayi1999,RezayiRead2009,BondersonSlingerland2008,BondersonFeguinMollerSlingerland2009,Hermanns,Sreejith11}, a consensus has not yet been achieved. (These states correspond to $\nicefrac{2}{5}$ and $\nicefrac{3}{5}$ in the second Landau level (LL), because including the spin degree of freedom, the lowest LL in GaAs corresponds to filling factor range $0<\nu<2$ and the second LL to $2<\nu<4$.) We report here on extensive tests of the so-called tripartite composite fermion (TCF) wave functions for these states. For the ground state, the TCF wave function is identical to the Read-Rezayi (RR) wave function \cite{ReadRezayi1999,CappelliGeorgievTodorov2001}, but the TCF form also enables a construction of the neutral and quasiparticle excitations, by creating excitations in individual partitions using the composite fermion (CF) theory \cite{Jain1989,Jainbook}. We will explore the validity of the TCF wave functions for a model interaction for which the RR wave function is exact, as well as for the Coulomb interaction.

As a brief background, the FQHE in the lowest LL at filling factors of the form $j\pm{n\over 2pn\pm 1}$ ($j$, $n$, $p$ integers) is understood as the integer quantum Hall effect (IQHE) of weakly interacting composite fermions carrying $2p$ vortices \cite{Jain1989,Jainbook}. 
(The Laughlin $\nicefrac{1}{m}$ states \cite{Laughlin1983} ($m$ odd), are seen as unit filling fraction of composite fermions carrying $m-1$ vortices.) 
The weaker FQHE states in the lowest LL at $\nicefrac{4}{11},\nicefrac{5}{13}, \nicefrac{5}{17} $, and $\nicefrac{6}{17}$ \cite{Pan03} are believed to arise from a \emph{fractional} QHE of interacting composite fermions, \cite{Chang04} although their precise nature is not yet fully established. The even denominator fraction $\nicefrac{5}{2}$ in the second LL \cite{observationofEvenDen,PanXiafivehalf1999} can also not be understood in terms of weakly interacting composite fermions, which would have produced a compressible CF Fermi sea here as in the lowest LL \cite{HalperinLeeRead1993}.
The most likely candidate for the $\nicefrac{5}{2}$ FQHE is the Moore-Read (MR) wave function, which was motivated by a conformal field theory construction \cite{MooreRead1991} and describes 
a chiral p-wave pairing of composite fermions  \cite{ReadGreen2000}. The FQHE at $\nicefrac{3}{8}$ \cite{Pan03,Bellani} and $2+\nicefrac{3}{8}$\cite{PanXia2004,PanXia2008,Kang} might also arise from CF pairing.\cite{TokeShiJain2008,LeeScarolaJain2002,ScarolaJainRezayi2002,MukherjeeMandalWojs2012}
New physics also seems possible for the second LL FQHE states  \cite{Kang,PanXia2008,PanXiafivehalf1999,PanXia2004} at $2+\nicefrac{1}{3}$, $2+\nicefrac{2}{3}$, $2+\nicefrac{3}{5}$ and $2+\nicefrac{2}{5}$, even though these nominally belong to the CF-IQHE sequence $2+\nicefrac{n}{(2n\pm 1)}$.  A quantitative investigation of the second LL fractions $2+\nicefrac{1}{3},2+\nicefrac{2}{3}$ shows substantial deviations from the non-interacting CF theory for the ground state as well as excitations \cite{MorfAmbrumenil1995,Ajit2012}, but is nonetheless likely (though not yet fully proven) that these states are adiabatically related to those at \nicefrac{1}{3} and \nicefrac{2}{3} in the lowest LL, although strongly renormalized by inter-CF interaction \cite{Ajit2012}. The FQHE states at $2+\nicefrac{2}{5}$ and $2+\nicefrac{3}{5}$ appear very different from the lowest LL states at $\nicefrac{2}{5}$ and $\nicefrac{3}{5}$, at least for small systems where exact results are available (see below), suggesting the possibility of a new mechanism for FQHE at these fractions. If $2+\nicefrac{2}{5}$ and $2+\nicefrac{3}{5}$ are not CF-IQHE states, then the recently observed fraction \cite{KumarCsathyManfra2010} $2+\nicefrac{6}{13}$ is also unlikely to be a CF-IQHE state.

\begin{figure}
\includegraphics[scale=.32]{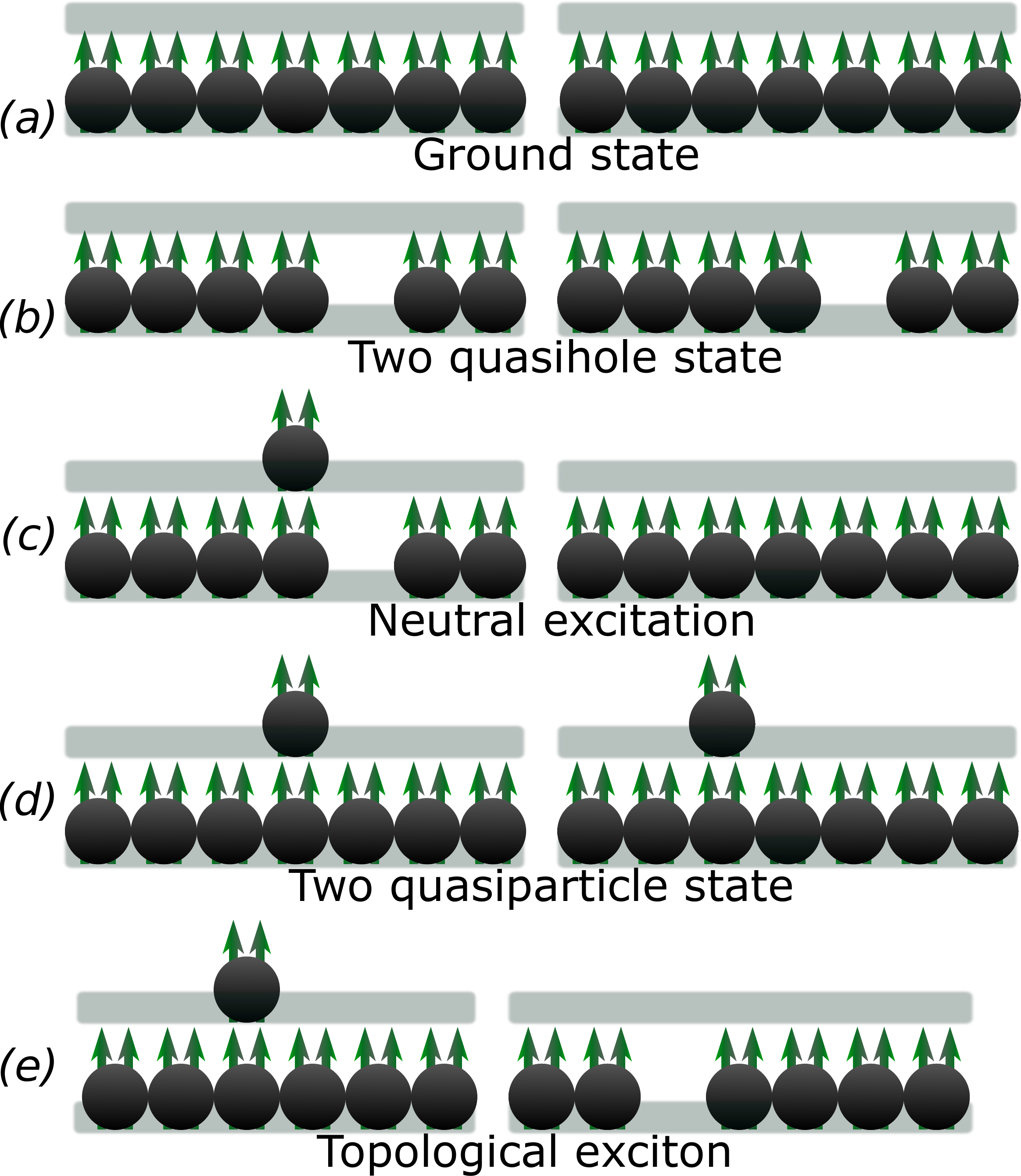}
\caption{Schematic description of the bipartite-CF construction for the $\nu=2+\nicefrac{1}{2}$ FQHE. The horizontal lines represent $\Lambda$ levels and the composite fermions are depicted as electrons bound to two flux quanta (or vortices). Each panel shows the distribution of composite fermions in the $\Lambda$Ls of the two partitions. Incompressible state (a) has lowest $\Lambda$L completely filled. Quasihole (b), quasiparticle (d) and neutral excitations (c) states have corresponding excitations in the individual partitions. A state with odd number of particles resembles an excitonic state in which the quasiparticle and quasihole are in separate partitions, as shown in panel (e).}
\label{BCF}
\end{figure}

The origin of FQHE at $2+\nicefrac{2}{5}$ and $2+\nicefrac{3}{5}$, which are related by particle hole symmetry (in the absence of LL mixing), is the subject of the present paper. Given that the $2+\nicefrac{1}{3}$ and $2+\nicefrac{2}{3}$ states are very likely adiabatically connected to the IQHE of composite fermions, and that $2+\nicefrac{1}{2}$ is likely to be a paired state of composite fermions, it is natural to suspect that the states at the intermediate fillings $2+\nicefrac{2}{5}$ and $2+\nicefrac{3}{5}$ are also described in terms of interacting composite fermions.
We study below a tripartite construction that builds interactions between composite fermions in a fashion motivated by the Read-Rezayi proposal \cite{ReadRezayi1999} for the $2+\nicefrac{3}{5}$ FQHE state, originally motivated by a conformal field theory construction. To introduce the idea of multipartite composite fermions, we begin with the bipartite composite fermion (BCF) representation \cite{Hermanns,Sreejith11,RodriguesSterdyniakHermannsSlingerland2012,CappelliGeorgievTodorov1999} of the MR Pfaffian wave function \cite{MooreRead1991} for the $2+\nicefrac{1}{2}$ FQHE, which represents a paired state of composite fermions. A general BCF wave function for $N=N_1+N_2$ composite fermions is written as (suppressing ubiquitous Gaussian factors)
\begin{equation}
\Psi_{\rm{BCF}}\sim\mathcal{A}\left[ \psi_{\nu_1}^{\rm CF}(z)\psi_{\nu_2}^{\rm CF}(w)\prod_{i}^{N_1}\prod_{j}^{N_2}(z_i-w_j) \right],
\end{equation}
where composite fermions are divided into two partitions $\{z_j\}$ and $\{w_j\}$ which have $N_1$ and $N_2$ particles respectively; they occupy the CF states $\psi_{\nu_1}^{\rm CF}$ and $\psi_{\nu_2}^{\rm CF}$ at fillings $\nu_1$ and $\nu_2$; and $z$ and $w$ denote the positions of particles as complex numbers of the form $x-iy$. The composite fermions in different partitions are inter-correlated through the cross term $\prod_{i}^{N_1}\prod_{j}^{N_2}(z_i-w_j)$. The operator $\mathcal{A}$ antisymmetrizes the entire wave function to generate a valid fermionic wave function. (Without the antisymmetrization, this wave function applies to a bilayer system \cite{HalperinHelPhysAct1983,ScarolaJain2001}) The MR wave function is reproduced when composite fermions in both partitions fill one $\Lambda$ level ($\Lambda$L), i.e. the corresponding electrons form the Laughlin $\nicefrac{1}{3}$ state, as shown schematically in figure \ref{BCF}(a). A nontrivial advantage of this representation is that excitations can be constructed by creating excitations within the individual partitions in the standard fashion, by exciting composite fermions to higher $\Lambda$Ls, as shown in fig.\ref{BCF}(b,c,d) for charged and neutral excitations. Furthermore, it provides an insight into the structure of an unpaired composite fermion, which is present for an odd number of particles. This state has been shown to contain a ``topological'' exciton~\cite{Hansson2009} (Fig.\ref{BCF}e) whose quasiparticle and quasihole cannot recombine~\cite{UnpairedCF_Bipartite}. Numerical studies in finite size systems show excellent agreement between the BCF model and the low energy spectra for a three-body interaction model, and also show adiabatic continuity to the Coulomb solution for the MR ground state~\cite{MorfDasSarma2010} and the topological exciton \cite{Sreejith11,UnpairedCF_Bipartite} (although such adiabatic connection has not been established for charged excitations in small system studies \cite{Toke07}). 

In the same spirit, tripartite composite fermion (TCF) states for $N=N_1+N_2+N_3$ particles are constructed as 
\begin{eqnarray}
&&\Psi_{\rm{TCF}} \sim \mathcal{A}\left[\psi_{\nu_1}^{\rm CF}(\mathbf{z})\psi_{\nu_2}^{\rm CF}(\mathbf{w})\psi_{\nu_3}^{\rm CF}(\mathbf{r})\times \vphantom{\prod_{i=1}^{N_1}\prod_{j=1}^{N_2}(z_i-w_j) \prod_{k=1}^{N_1}\prod_{l=1}^{N_3}(z_k-r_l) \prod_{p=1}^{N_2}\prod_{q=1}^{N_3} (w_p-r_q)} \right. \nonumber \\
&& \left.\vphantom{\psi(z)\psi(w)\psi(r)} \prod_{i=1}^{N_1}\prod_{j=1}^{N_2}(z_i-w_j) \prod_{k=1}^{N_1}\prod_{l=1}^{N_3}(z_k-r_l) \prod_{p=1}^{N_2}\prod_{q=1}^{N_3} (w_p-r_q) \right]
\end{eqnarray}
where $\mathbf{w}=\{w_1,w_2\dots\}$, $\mathbf{z}=\{z_1,z_2\dots\}$ and $\mathbf{r}=\{r_1,r_2\dots\}$ denote an arbitrary partition of the $N$ particles into three parts with $N_1$, $N_2$ and $N_3$ particles respectively, and $\psi_{\nu_1}^{\rm CF}$, $\psi_{\nu_2}^{\rm CF}$ and $\psi_{\nu_3}^{\rm CF}$ are CF states formed within the individual partitions. The TCF state in which the composite fermions completely fill the lowest $\Lambda$L in each partition is an exact representation of the $k=3$ RR state ($\mathbb{Z}_3$ parafermion state) at $\nu=\nicefrac{3}{5}$. This state was first proposed as a generalization of the Pfaffian wave function \cite{ReadRezayi1999} and has attracted interest because of the possibility of its particle-hole conjugate being realized at $\nu=2+\nicefrac{2}{5}$, and also because 
of the possibility of this state supporting excitations with non-trivial braiding properties.

The RR state is the maximum density exact ground state of a 4-body model Hamiltonian $\mathcal{H}_4$ (described in section II). Wave functions can also be constructed for quasiholes of this state which are also exact solutions of the same model Hamiltonian. Explicit wave functions of such quasihole states as well as their counting were given in Refs.~[\onlinecite{Read2006RRQH,ArdonneSchoutensQHofRR2007,ArdonneKedemStone}]. However, as is true for all truncated pseudopotential models, explicit solutions with {\em nonzero} energies, such as quasiparticles and neutral excitations, are not known. Similarly wave functions for states for which $N$ is not a multiple of $3$ are not known except when they contain only quasihole excitations. The TCF representation gives a natural way of constructing arbitrary excitations of $2+\nicefrac{3}{5}$ by exciting composite fermions in the individual partitions, and also wave functions for systems in which the total number of particles is not a multiple of three.

We make a number of simplifying assumptions in our analysis below. We assume that the magnetic field is large enough that all electrons are fully polarized; the spin degree of freedom is thus frozen and not considered explicitly. Landau level mixing is neglected, and the electrons in the lowest LL are treated as inert. The actual two dimensional electron systems used in experimental situations have a finite width, which in general produces a weakening of Coulomb repulsion at short distances. We have ignored such effects, because, as seen below, our current quantitative understanding of the $2+\nicefrac{2}{5}$ or $2+\nicefrac{3}{5}$ states is not at a level where inclusion of such corrections would be meaningful. Finally, we also neglect effects of disorder, which is always present and is known to substantially diminish the activation gaps.

The remainder of the article is organized as follows. The construction of the TCF wave functions is described in section \ref{sec:construction}. Section \ref{sec:numericalTechniques} summarizes the numerical techniques used in the various calculations presented in this work. Section \ref{sec:comparewith4BI} tests how the TCF wave functions compare with the low-energy solutions of the $\mathcal{H}_4$ interaction, for which the RR state is an exact solution, for systems with up to 16 particles. Section \ref{sec:comparewithCoulombGS} compares the RR ground state with the exact Coulomb ground state, and Section \ref{sec:TCF-CoulombExcitations} compares the TCF model for excitations with the exact Coulomb excitations. These sections also test if the solutions of the four-body interaction are adiabatically connected to those of the Coulomb interaction. The article is concluded in Section \ref{sec:Conclusion}. The appendix contains a multipartite generalization of the bipartite and tripartite CF wave functions.

Throughout this article, the phrase ``Coulomb interaction" should be taken to mean the ``{\em second LL} Coulomb interaction"  corresponding to the Coulomb interaction acting on the Hilbert space of the second LL wavefunctions.

\section{Construction of tripartite wave functions}
\label{sec:construction}

In this section, we describe the construction of general TCF wave functions of $N$ particles. Electron coordinates are represented as complex numbers $\{z_1,\ldots,z_N\}$ representing positions on the complex plane. Wave functions given below should to be multiplied by an overall geometric factor ($\prod_{i=1}^N \exp(-\frac{|z_i|^2}{4})$ for disk geometry) which has been omitted for brevity. 
The actual single particle wave functions in the second LL are functions of $z$ and $\bar{z}$. However, LL lowering operators bijectively map these states into the lowest LL wave functions which are functions of $z$ alone. For this reason, all the wave functions are written in the lowest LL. Difference in the action of Coulomb interaction on the states of the first and second LLs is accounted for by using the second LL Coulomb pseudopotentials.

\subsection{Incompressible TCF states}

Incompressible TCF states are defined to be those states in which each partition contains an incompressible state with fully occupied $\Lambda$Ls. A state with an equal number of particles in each partition, $N_1=N_2=N_3\equiv \widetilde{N}$, can be written as
\begin{align}
\Psi_\nu^{\rm{TCF}}=\mathcal{A} \left[ \psi_{\frac{n}{2pn+1}} (\mathbf{z}) \psi_{\frac{n}{2pn+1}} (\mathbf{w}) \psi_{\frac{n}{2pn+1}} (\mathbf{r})  \vphantom{\prod_{i,j=1}^{\widetilde{N}} (z_i-r_j)}  \right.\times   \nonumber \\
\left. \prod_{i,j=1}^{\widetilde{N}}(z_i-w_j) \prod_{k,l=1}^{\widetilde{N}} (z_k-r_l) \prod_{p,q=1}^{\widetilde{N}} (w_p-r_q) \right],
\label{TCFIncompressible}
\end{align}
where $\mathbf{w}=\{w_1,w_2\dots,w_{\widetilde{N}}\}$, $\mathbf{z}=\{z_1,z_2\dots,z_{\widetilde{N}}\}$ and $\mathbf{r}=\{r_1,r_2\dots,r_{\widetilde{N}}\}$ is an arbitrary partition of the $N=3\widetilde{N}$ particles into three equal parts. The particles in different partitions are correlated through the cross terms, whereas those in each individual partition form the Jain CF state at filling fraction $\bar{\nu}=\frac{n}{2pn+1}$, given by 
$
\psi_{\frac{n}{2pn+1}} =\mathcal{P}_{\rm LLL} \prod_{i<j=1}^{\widetilde{N}} (z_i-z_j)^{2p} \Phi_{n}
$
where $\Phi_{n}$ is the wave function of an integer quantum Hall system in which $n$ Landau levels are completely filled, $p$ is an integer, and $\mathcal{P}_{\rm LLL}$ projects the state into the lowest LL. 
The one filled $\Lambda$L state $\psi_{\frac{1}{3}}$ reduces to the Laughlin wave function \cite{Laughlin1983}.

The filling factor of the TCF wave function can be derived by noting that the largest power of an electron coordinate, which is the angular momentum of the outermost occupied single particle state in the disk geometry, is $L_{\rm max}=\nicefrac{(1+2pn+2n)N}{n}$, where $\Psi_{\bar{\nu}}$ contributes $\nicefrac{N}{\bar{\nu}}+\mathcal{O}(1)\approx\nicefrac{(2pn+1)N}{n}$ and the cross term contributes $2{\widetilde{N}}$. For a total of $N=3{\widetilde{N}}$ particles, the overall filling fraction of the TCF wave function is 
\begin{equation}
\nu=\frac{3\widetilde{N}}{L_{\rm max}}=\frac{3\bar{\nu}}{1+2\bar{\nu}}=\frac{3n}{1+(2p+2)n}.
\end{equation}
In this paper we study only the $\nu=\nicefrac{3}{5}$ TCF functions in which individual partitions have a filling fraction $\bar{\nu}=\nicefrac{1}{3}$. Unless otherwise stated, the phrase ``TCF wave function" refers to $\nu=\nicefrac{3}{5}$ TCF wave function in what follows.

\begin{figure}
\includegraphics[scale=.32]{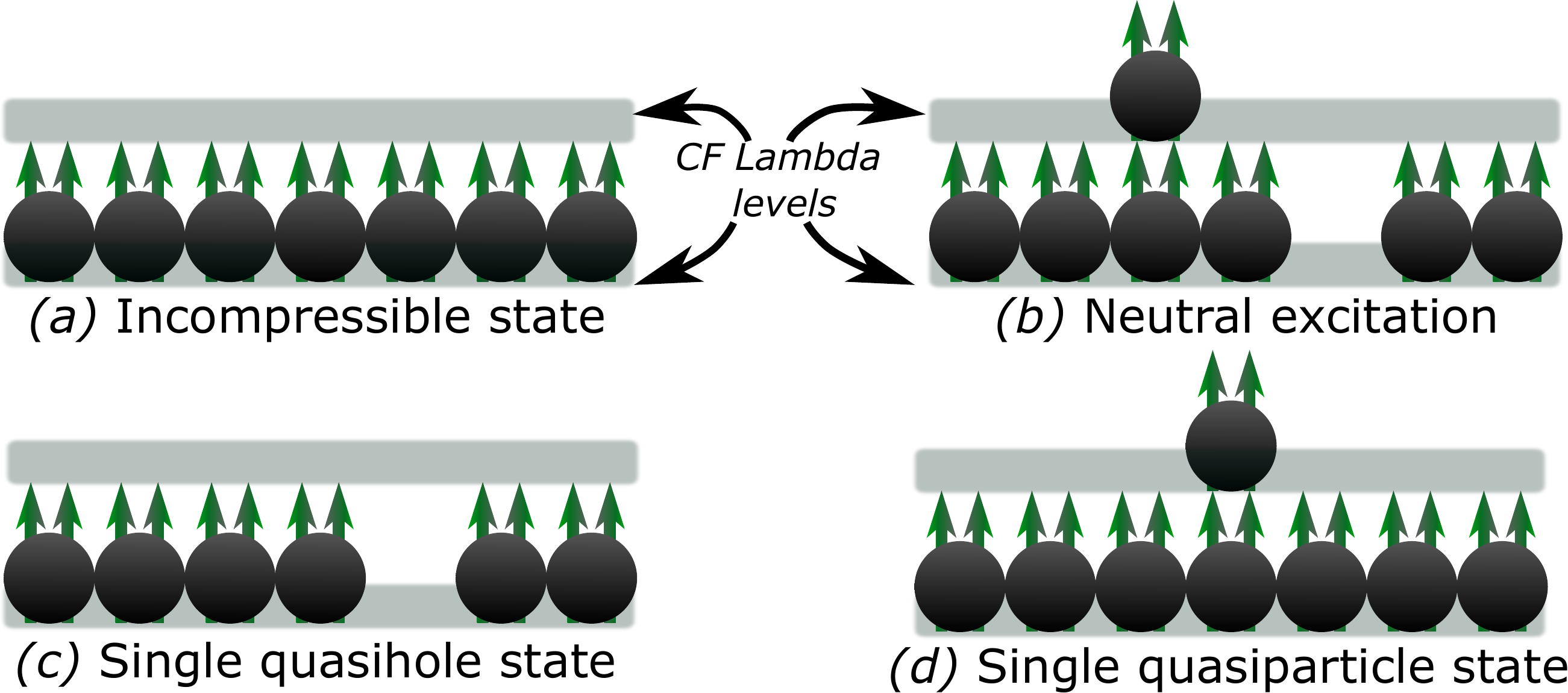}
\caption{Schematic description of the incompressible state and its charged and neutral excitations.  Panel (a) shows the 1/3 incompressible state which has one fully occupied $\Lambda$ level, while the panels (b), (c) and (d) show neutral excitation (a particle hole pair of composite fermions), a quasihole (a missing composite fermion), and a quasiparticle (an additional composite fermion), respectively.}
\label{CFdescription}
\end{figure}

\subsection{Construction of TCF states on a sphere}

The bulk properties of a quantum Hall system, which are of interest in this work, are most conveniently studied in the spherical geometry~\cite{Haldane1983}, primarily due to the absence of edges. In this model, the electrons move on the surface of a sphere in the presence of a uniform and constant radial magnetic field generated by a monopole placed at the center of the sphere. The strength of the monopole is represented by $Q$ which is defined as half the number of total magnetic flux quanta emitted by the monopole. Single-valuedness of electronic wave functions requires that the monopole strength $Q$ be an integer or a half integer. Landau quantization causes formation of discrete electronic kinetic energy levels. The states of the $n^{\rm th}$ Landau level form a multiplet of total angular momentum quantum number $l=Q+n$, where $n=0,1,2,\dots$. This gives a total degeneracy of $2(Q+n)+1$ to the $n^{\rm th}$ LL.

In particular, the lowest LL orbitals have angular momentum quantum number $Q$, and the single particle state with $z$ component $-Q\leq m\leq Q$ is given by $u^{Q+m}v^{Q-m}$, where $u=\cos(\theta/2)e^{i\phi/2}$ and $v=\sin(\theta/2)e^{-i\phi/2}$ are the spinor coordinates on the sphere. The TCF wave function (equation \ref{TCFIncompressible}) can be transcribed straightforwardly to the spherical geometry; in particular, the factors $(z_i-z_j)$ are replaced by $(u_iv_j-u_jv_i)$.

The monopole strength at which a given wave function occurs can be inferred by noting that the largest power of $u_i$ gives $2Q$. 
Applying this to the wave function of the incompressible TCF wave function tells us that it occurs at a flux
\begin{equation}
2Q = \frac{5}{3}N-3.
\label{IncompressibleTCFFlux}
\end{equation}
The fact that incompressible states can occur only for particle numbers that are multiples of three ensures that $2Q$ is an integer. In general, a quantum Hall trial wave function at filling fraction $\nu$ occurs at a flux given by 
\begin{equation}
2Q=\frac{1}{\nu}N-s
\end{equation}
where $s$ is called the shift. The shift for the RR wave function is $3$.

\subsection{Jack polynomials}

The RR states and their quasihole excitations (including the MR Pfaffian) have been identified as Jack polynomials~\cite{Bernevig2008}, which allow efficient numerical generation of these states. We briefly review this method here and will use it below to calculate the overlap between RR ground state and Coulomb eigenstate. The single-particle states in the lowest LL (LLL) are indexed by their angular momenta and there are two equivalent ways of representing a non-interacting $N$-particle state. One can label it by a partition (not to be confused with the word ``partition" used in describing ``tripartite composite fermion") $\lambda=\left[\lambda_1,\cdots,\lambda_N\right]$ in which the occupied single-particle angular momenta are listed with $\lambda_1\geq \lambda_2 \cdots \geq \lambda_N$. Or one can list the occupation number of orbitals as $n=\{n_m\}$, $m=0,1,2,\cdots$, where $n_m$ is the number of particles in the orbital labeled by $m$. A non-interacting many-body wave function is a Slater determinant which can be labeled by a partition or an occupation. An interacting many-body state is a superposition of many non-interacting basis states indexed by $\lambda$'s with coefficients $c_\lambda$. The squeezing operation for partitions is defined as follows: for a pair of particles in the orbitals $m_1$ and $m_2$, with $m_1<m_2-1$, the elementary squeezing operation corresponds to shifting two particles inwards by moving a particle each from orbital $m_1$ to $m_1+1$ and from orbital $m_2$ to $m_2-1$. Equivalently in terms of occupation numbers, squeezing decreases $n_{m_{1,2}}$ by one and increases $n_{m_1+1}$ and $n_{m_2-1}$ by one. A partition $\lambda$ is said to dominate $\mu$ (denoted as $\lambda>\mu$) if $\mu$ can be generated by squeezing $\lambda$. A fermionic Jack can be expanded in terms of Slater determinants
\begin{equation}
J_{\lambda}^{\alpha}=\sum_{\kappa\le\lambda}c_{\lambda\kappa}(\alpha){\rm sl}_\kappa,
\end{equation}
where $\alpha$ is a parameter, the sum over $\kappa$ runs over all partitions squeezed from the root partition $\lambda$ and ${\rm sl}_\kappa$ is the Slater determinant labeled by $\kappa$. There is a recursive relation~\cite{Thomale2011} for the expansion coefficients $ c_{\lambda \kappa}(\alpha)$
\begin{equation}
c_{\lambda \kappa}(\alpha)=\frac{2(1/\alpha-1)}{\rho_{\lambda}(\alpha)-\rho_{\kappa}(\alpha)} \sum_{\kappa<\mu\le\lambda}\hspace{-5pt}\left(l_i-l_j\right)c_{\mu \kappa}(\alpha)(-1)^{N_{\rm SW}},
\label{JackExpan}
\end{equation}
where the sum is over all partitions $\mu=\left[l_1,\cdots,l_i+s,\cdots,\l_j-s,\cdots,l_N\right]$ that strictly dominate $\kappa=\left[l_1,\cdots,l_i,\cdots,\l_j,\cdots,l_N\right]$ but being dominated or equal to the root partition $\lambda$. The $\rho$'s are defined as:
\begin{equation}
\rho_{\lambda}(\alpha)=\sum_i \lambda_i \left(\lambda_i + 2i(1-1/\alpha)\right).
\end{equation}
The quantity $N_{\rm SW}$ is the number of swaps that are needed to bring $\mu$ back to $\kappa$. For the RR $Z_3$ state, the root occupation is $1110011100\cdots00111$ and the parameter $\alpha$ is $-4$.

\subsection{TCF excitations}

The structure of the incompressible TCF wave function suggests a natural way of constructing excitations by introducing neutral or charged excitations within the individual partitions. Figure \ref{CFdescription} schematically shows the $\Lambda$L occupation of composite fermions for the incompressible state and for various excitations at $\nu=\nicefrac{1}{3}$. The lowest energy neutral excitations of an incompressible TCF state are obtained by creating the lowest energy neutral excitation in \emph{one} of the partitions of the TCF wave function. Charged excitations are obtained by changing the flux by one unit. Addition of a flux quantum to the incompressible system results in a wave function wherein there is one quasihole in each of the three partitions. Removal of a flux similarly results in one quasiparticle in each partition. For these states, there are equal number of particles in each partitions, just as in the case of the incompressible state.

\begin{figure}
\includegraphics[scale=.45]{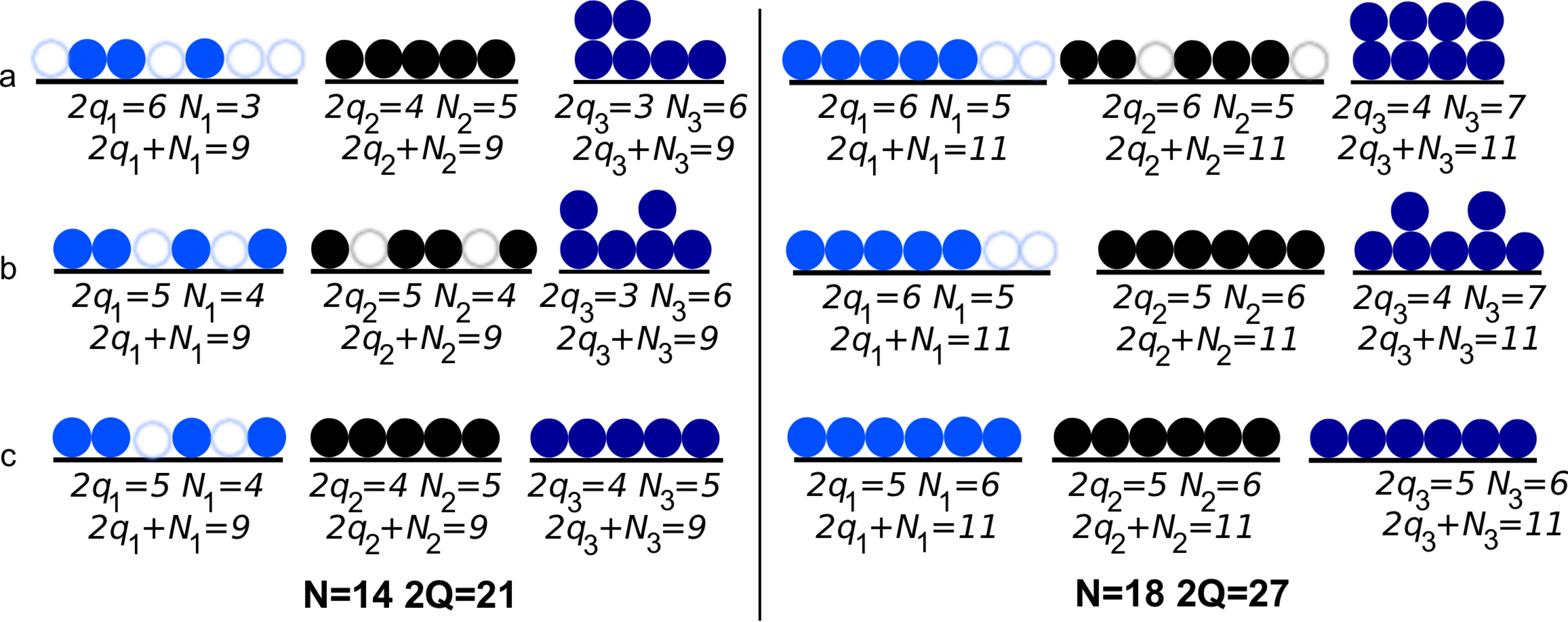}
\caption{For a given $N$ and $2Q$ (total number of particles and total flux), there are several wave functions that satisfy the conditions in equation \ref{constraint}. The figure schematically shows different possible wave functions for two cases, $(N,2Q)=(14,21)$ and $(18,27)$. The composite fermions in different partitions are shown by different colors, and their arrows have been suppressed to avoid clutter.  The value of $N_i$, $q_i$ and $N_i+2q_i$ are given below the individual partitions. Case (c) has the lowest total CF cyclotron energies in both examples.} \label{listoptions}
\end{figure}

\begin{figure*}
\includegraphics[width=.9\textwidth]{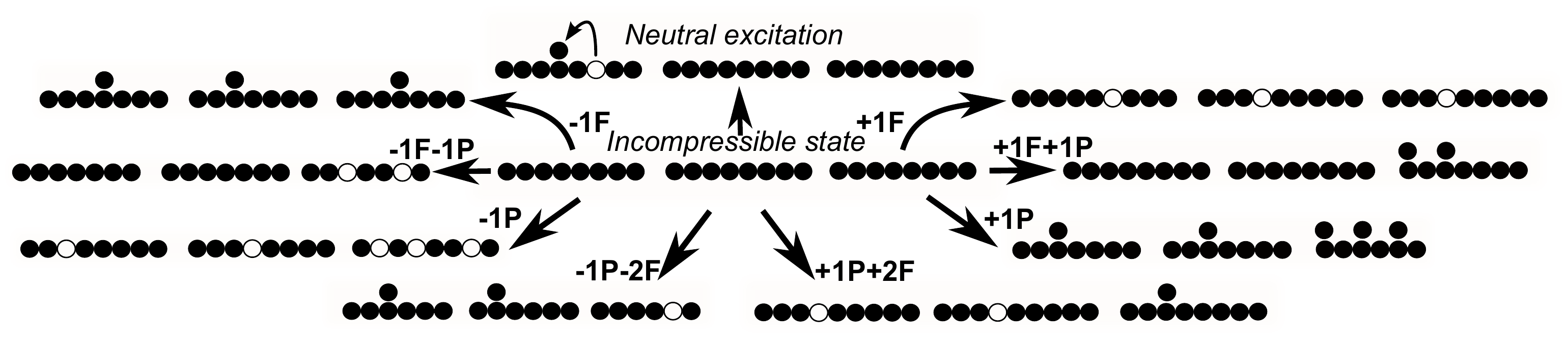}
\caption{Schematic depiction of various TCF states.  The incompressible state (center) has the composite fermions fully occupying the lowest $\Lambda$ level in each partition. (The arrows of composite fermions have been suppressed for simplicity.)
Excited states are obtained by either creating an excitation in one of the partitions (top) or by addition or removal of a flux and / or an electron. The symbols $+1F$ ($-1F$) and $+1P$ ($-1P$) represent addition (removal) of a flux and electron respectively.}\label{schematicTCF}
\end{figure*}

We can also consider states for which the particle number $N$ is not a multiple of three, so the numbers of particles in the partitions are not equal. Consider a wave function in which the partitions contain $N_1$, $N_2$ and $N_3$ electrons, and the effective flux experienced by the composite fermions in the individual partitions be $q_1$, $q_2$ and $q_3$ respectively. After including the contributions from Jastrow factors ($N_i-1$ in $i$th partition) and the cross terms ($N_j+N_k$; $j,k\neq i$), the net flux experienced by the electrons in the three partitions are 
\begin{eqnarray}
\nonumber 2Q_1 = 2q_1 + 2(N_1-1) + (N_2+N_3) \\ 
\nonumber 2Q_2 = 2q_2 + 2(N_2-1) + (N_1+N_3) \\
2Q_3 = 2q_3 + 2(N_3-1) + (N_1+N_2).
\end{eqnarray}
Because all the electrons must ultimately reside in the same Hilbert space in the fully antisymmetrized wave function, the total fluxes experienced by the electrons should be identical, i.e. $2Q_1=2Q_2=2Q_3\equiv 2Q$, which implies the constraint 
\begin{equation}
N_i+2q_i = \text{constant } 2Q + 2 - N, \text{ for } i=1,2,3.
\label{constraint}
\end{equation}
For given $N$ and $Q$, there are several wave functions that satisfy the above constraints. Figure \ref{listoptions} shows the possible wave functions that satisfy the constraints for two specific examples. It is natural to pick the state with the lowest total CF cyclotron energy as the trial wave function in each case.  Figure \ref{schematicTCF} shows the structure of the several simple TCF excited states. Note that, due to antisymmetrization, permutations of the three partitions do not give new wavefunction.

It is straightforward to determine the local charge excess or deficiency associated with a quasiparticle or quasihole. The adiabatic insertion (removal) of one flux quantum produces an overall charge equal to the filling factor. However, this corresponds to three quasiholes (quasiparticles), one in each partition. The charge of an elementary quasiparticle or quasihole thus has a magnitude of 
\be
\frac{e^*}{e}=\frac{\nu}{3}=\frac{n}{(2pn+2n+1)}.
\ee

\subsection{Angular momentum of the TCF states}

We derive the useful result that the total angular momentum $L_z$ of the TCF wave function is the sum of the $L_z$ of states in individual partitions. In terms of the spinor coordinates $(u_i,v_i)$, the angular momentum operator $L_z$ is given by
\begin{equation}
L_z=\sum_{i=1}^N \frac{1}{2}(u_i\partial_{u_i}-v_i\partial_{v_i})
\end{equation}
The operator commutes with the antisymmetrization operator. Furthermore, the action of this operator on the cross terms vanishes as a result of the identity $L_z(u_i v_j-v_iu_j)=0$. Therefore, we have 
\begin{eqnarray}
L_z \Psi_{\rm{TCF}}	&=& \mathcal{A}\left[ \{L_z\psi(z)\}\psi(w)\psi(r)\times\rm{cross}\,\rm{terms} \right] \nonumber\\
				  	&+&\mathcal{A}\left[ \psi(z)\{L_z\psi(w)\}\psi(r)\times\rm{cross}\,\rm{terms} \right] \nonumber\\
				  	&+&\mathcal{A}\left[ \psi(z)\psi(w)\{L_z\psi(r)\}\times\rm{cross}\,\rm{terms} \right] \nonumber\\
					&=&(L_z^{(1)}+L_z^{(2)}+L_z^{(3)}) \Psi_{\rm{TCF}}
\end{eqnarray}
The angular momenta of the state in each partition can be obtained by adding the $L_z$ quantum numbers of individual electrons. Alternatively, relative to the incompressible state, we can simply add the angular momenta of the excitations, with the angular momenta of quasiholes taken as negative.

\section{Numerical methods}
\label{sec:numericalTechniques}

\subsection{four-body model Hamiltonian}

The incompressible TCF wave function is the highest density exact zero energy state of a four-body model Hamiltonian \cite{ReadRezayi1999,Read2006RRQH,SimonRezayiCooper2007} which can be written as
\begin{equation}
\mathcal{H}_4=\sum_{i<j<k<l=1}^N \mathcal{P}_{ijkl}(4Q-6)
\label{FourBodyHamiltonian}
\end{equation}
The operator $\mathcal{P}_{ijkl}(M)$ projects a many-particle state to the total angular momentum $M$ sector of Hilbert space of the four particles $i,j,k,l$. Angular momentum $4Q-6$ is the largest total angular momentum sector in the four particle Hilbert space within the lowest Landau level. This corresponds to the closest approach of the four particles. Summation over all possible four particle sets gives a valid quantum mechanical Hamiltonian operator for which there is an energy cost for $4$ particles `approaching' each other at an angular momentum $4Q-6$ and none otherwise.

Numerically, the four body Hamiltonian on a sphere with monopole strength $Q$ is constructed by first obtaining the Clebsch Gordan coefficients $\mathcal{C}^{L,m}_{m_1,m_2,m_3,m_4}$ corresponding to states of total angular momentum $L=4Q-3$ through diagonalizing the $L^2$ operator. In terms of these coefficients, the four body interaction can be represented as
\begin{eqnarray}
\mathcal{H}_4&=&\sum_{\{m_i\},\{n_i\}} c^\dagger_{n_4}c^\dagger_{n_3}c^\dagger_{n_2}c^\dagger_{n_1} \chi^{\{n_i\}}_{\{m_i\}} c_{m_1}c_{m_2}c_{m_3}c_{m_4}\\
\chi^{\{n_i\}}_{\{m_i\}}&=& \mathcal{C}^{L,n_1+n_2+n_3+n_4}_{n_1,n_2,n_3,n_4}\mathcal{C}^{L,m_1+m_2+m_3+m_4}_{m_1,m_2,m_3,m_4}\nonumber
\end{eqnarray}
where $L=4Q-6$ and $c^\dagger_m$ is the electron creation operator for the state of $z$ component angular momentum $m$.

The quasihole excitations of the RR state can also be written as the exact zero energy states of $\mathcal{H}_4$.\cite{Read2006RRQH} However there are no zero energy states on the quasiparticle side.

\subsection{TCF states and spectra}

The finite systems studied here are realized in the spherical geometry. The TCF wave functions are obtained by antisymmetrization of the product of the Jain CF wave functions and the cross terms. Evaluation of the TCF wave function is computationally slow because each evaluation involves $N!/(N_1!N_2!N_3!)$ antisymmetrization steps, which renders Monte Carlo methods unfeasible for the evaluation of overlaps, energies etc. We have devised a method that employs the complete set of simultaneous eigenstates $L^2$ and $L_z$ on the sphere obtained by exact diagonalization.

To diagonalize the TCF states and to calculate their overlaps with the exact eigenstates, we need to construct TCF wave functions that are eigenstates of the angular momentum operators $L^2$ and $L_z$, which we obtain as follows. Consider the sector with total angular momentum quantum number $L=M$. Let $\{\Psi_1$, $\Psi_2,\dots,\Psi_K\}$ be the set of all \emph{linearly independent} TCF states with $z$ component angular momentum $L_z=M$. (Method for numerically obtaining a linearly independent set is described later in this section.)
By diagonalizing (using Lanczos algorithm) the angular momentum operator in the Hilbert space of all $L_z=M$ \emph{Slater} determinant states on the sphere, we can obtain all the $L=L_z=M$ eigenstates $\{\phi_1,\phi_2,\dots \phi_P\}$.
If there is a state with angular momentum $L=L_z=M$ in the TCF sector, then it should be possible to write that state in terms of the states $\phi_i$, since the later gives a complete basis. In other words, there should be a solution for $c_i$ and $d_i$ in the following equation
\begin{equation}
c_1 \Psi_1 + c_2 \Psi_2 +\dots + c_K\Psi_K = d_1 \phi_1 + d_2 \phi_2 \dots d_P \phi_P\nonumber
\end{equation}
In order to solve this, we use the fact that this equation is true for any configuration $\mathbf{z}$ of the electrons. The functions $\phi_i$ and $\Psi_i$ are evaluated at a large number of randomly obtained configurations $\mathbf{z}_1,\mathbf{z}_2,\dots$ giving a sufficiently large linear system of equations
\begin{eqnarray}
c_1 \Psi_1(\mathbf{z}_i) + c_2 \Psi_2(\mathbf{z}_i) +\dots + c_K\Psi_K(\mathbf{z}_i) =\nonumber\\
= d_1 \phi_1(\mathbf{z}_i) + d_2 \phi_2(\mathbf{z}_i) \dots d_P \phi_P(\mathbf{z}_i) \nonumber
\end{eqnarray}
where $i$ indexes the different configurations.
There are as many independent solutions to the above set of equations as there are $L=L_z=M$ states in the TCF space. 

Once the $L$-$L_z$ eigenstates are constructed, it is straightforward to diagonalize a given Hamiltonian within the TCF basis to obtain the TCF eigenstates and eigenenergies. Given the coefficients $d_i$ and the Slater determinant expansions for $\phi_i$ it is then straightforward to expand the TCF state itself in the Slater determinant states. Energy of the Slater determinant expansion is obtained by using the pseudopotentials for the interaction of interest. While this method in principle gives exact results, it is most efficient if the functions are scaled such that $\Psi_i(z)$ and $\phi_i(z)$ have similar orders of magnitude.

\subsection{Identification of linearly independent trial states}

The set of all TCF states with a particular $L_z$ can be constructed by selecting those arrangements of excitations that result in the desired $L_z$. However, such a set is in general not linearly independent. Schemes to generate linearly independent quasihole states of general $n$ body Hamiltonians exist\cite{ArdonneKedemStone,ArdonneSchoutensQHofRR2007,Read2006RRQH,ReadRezayi1996}. In finite systems, linearly independent states can be easily numerically identified for \emph{arbitrary} set of states.
If the finite set of functions $X=\{\phi_1,\ldots,\phi_S\}$ is linearly independent, there should be a non-trivial solution for $d_i$ in the equation
\begin{equation}
\sum_{i=1}^S d_i \phi_i = 0
\end{equation}
By evaluating the above statement for a large number of randomly chosen configurations $\mathbf{z}_j$, we get a set of simultaneous linear equations, which have as many solutions as there are linearly dependent states in the set. The number of such linear dependencies can be equivalently obtained by finding zeros in the singular valued decomposition of the matrix $A_{ij}=\phi_i(\mathbf{z}_j)$ where $i=1,2,\dots,S$ and $j>S$. By removing an appropriate number of states from the set $X$, one can obtain a linearly independent subset of $X$.

\subsection{Angular momentum counting of TCF states}

Number of angular momentum multiplets that can be constructed using the trial states in each \emph{total} angular momentum sector can be calculated by counting the number of highest weight vectors. If there are $k$ and $p$ linearly independent trial states with $z$-component angular momentum $m$ and $m+1$, then the number of highest weight vectors in the total angular momentum sector $L=L_z=m$ is $k-p$. For example, if the number of trial states with $z$-component angular momenta $(0,1,2,3,\ldots,M,M+1)$ are $(a_0,a_1,a_2,\ldots,a_M,0)$, then the number of states of total angular momentum quantum numbers $(0,1,2,\ldots,M,M+1)$ is $(a_0-a_1,a_1-a_2,\ldots,a_M-0,0)$. This method relies on the fact that the TCF space contains complete multiplets, in other words, if $\phi$ is a TCF state, then $L_- \Psi$ and $L_+ \Psi$ are also TCF states or 0. This is because $L_{\pm}=\sum_{i=1}^N \left[ L_\pm \right]_i$ commmutes with antisymmetrization operation as well as cross terms allowing one to write the action of the operators as 
\begin{eqnarray}
L_\pm \Psi_{\rm{TCF}}	&=& \mathcal{A}\left[ \{L_\pm \psi(z)\}\psi(w)\psi(r)\times\rm{cross}\,\rm{terms} \right] \nonumber\\
				  	&+&\mathcal{A}\left[ \psi(z)\{L_\pm \psi(w)\}\psi(r)\times\rm{cross}\,\rm{terms} \right] \nonumber\\
				  	&+&\mathcal{A}\left[ \psi(z)\psi(w)\{L_\pm \psi(r)\}\times\rm{cross}\,\rm{terms} \right] \nonumber\\
\end{eqnarray}
Action of $L_\pm$ on CF wavefunctions $\psi$ gives another CF state. Therefore, each of the three terms in the right hand side of above equation is a TCF wavefunction. Thus $L_\pm \Psi_{\rm{TCF}}$ is contained in the space of TCF wavefunctions.

\section{Comparison with exact spectrum of the four-body interaction}
\label{sec:comparewith4BI}

\begin{figure}
\includegraphics[scale=.58]{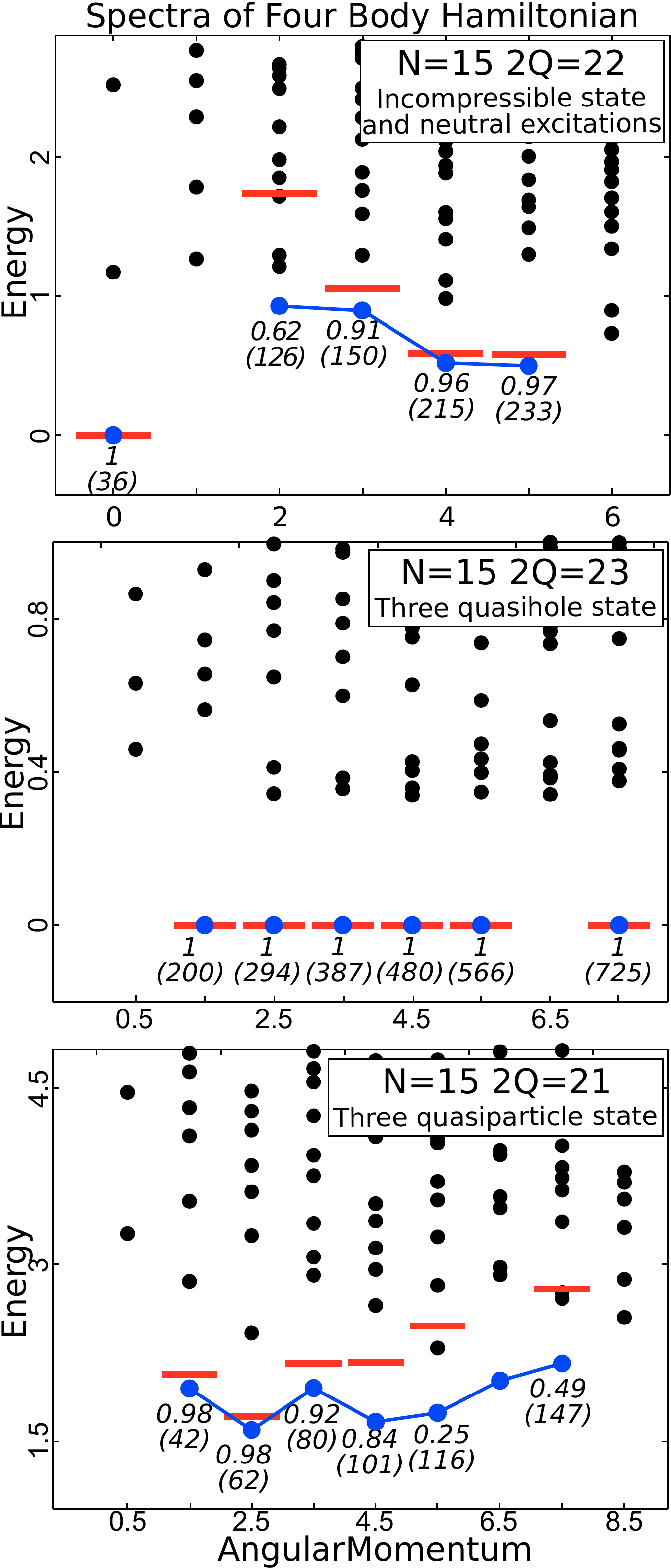}
\caption{Energy spectra of $N=15$ particles for the four-body model Hamiltonian ${\cal H}_4$ at flux values $2Q=21$, 22 and 23. The blue and black dots show the exact spectra (the blue dots mark the low energy states). The red dashes show the spectra evaluated within the TCF basis; the top panel shows the TCF spectrum for the $\nu=3/5$ ground state (also the RR state) and its neutral excitation; the middle and bottom panels show TCF spectra for three quasiholes and quasiparticles, respectively. The total angular momentum quantum number $L$ is shown on the $x$-axis and the energy on $y$-axis. The overlaps of the TCF states with the corresponding exact states are shown near each state; the parentheses contain the total numbers of independent eigenstates in the corresponding angular momentum sector.}
\label{Trial-4B-3N}
\end{figure}

\begin{figure}
\includegraphics[scale=.54]{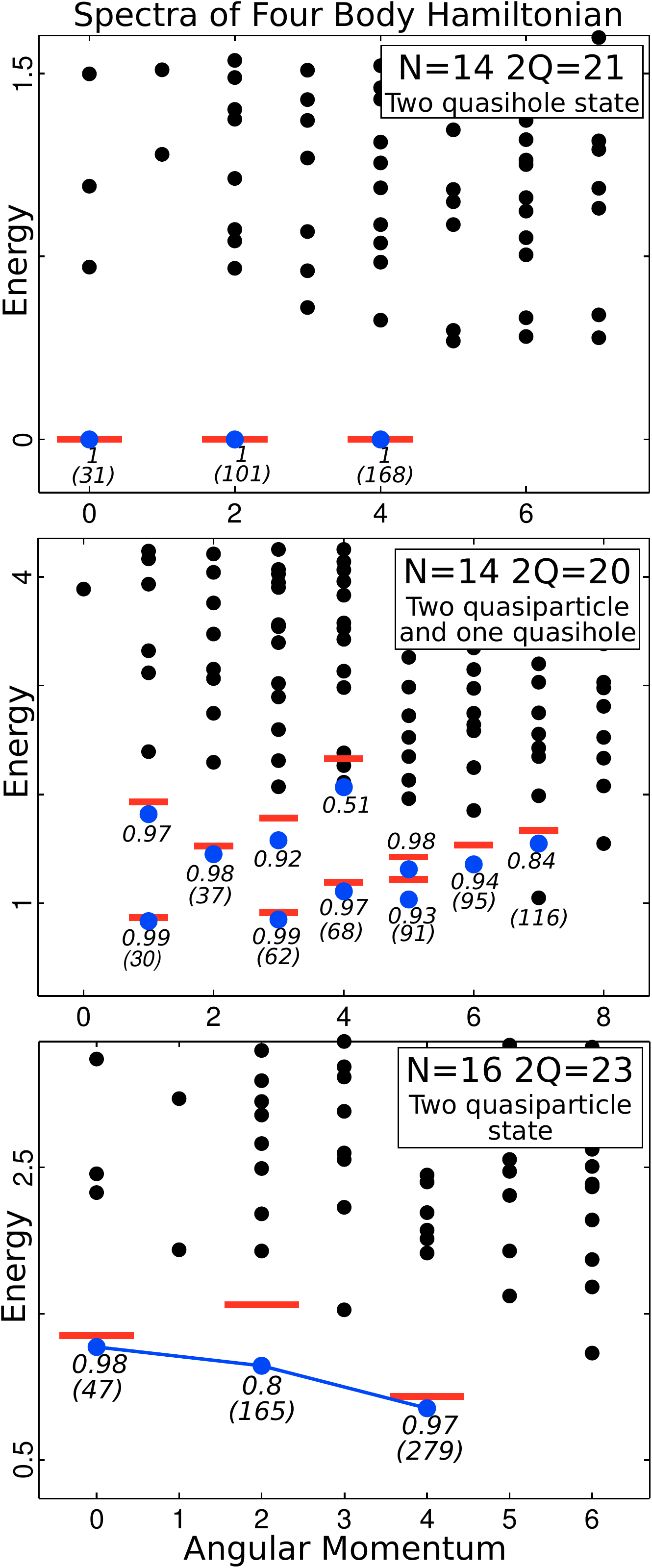}
\caption{Exact and TCF spectra for the four-body model Hamiltonian for $N=14$ particles at $2Q=20,$ 21 and 23. The meanings of various symbols are same as those in figure \ref{Trial-4B-3N}. These states are obtained from the incompressible state with $(N,2Q)=(15,22)$ (top panel of Fig.~\ref{Trial-4B-3N}) by addition or removal of electrons and fluxes; removal of one particle and one flux results in a 2 quasihole state (top); removal of one particle and two fluxes result in a state with two quasiparticles and one quasihole in different partitions (middle); addition of one particle and one flux results in a two quasiparticle state (bottom).}
\label{Trial-4B-3Npm}
\end{figure}

In this section, we diagonalize the four-body Hamiltonian ${\cal H}_4$ (i) within the TCF sector and (ii) within the full LLL Hilbert space, and compare the two sets of eigenenergies and eigenfunctions.  This will tell us to what extent the TCF states capture the low energy physics of the four-body interaction. The results from these comparisons are shown in figures \ref{Trial-4B-3N} and \ref{Trial-4B-3Npm} for systems sizes of up to $N=16$. The TCF spectra are shown by red dashes and the full exact spectra by black or blue dots; the overlaps between the TCF eigenstates and the corresponding exact eigenstates are also shown, along with the number of independent states in that sector. 

Figure \ref{Trial-4B-3N} shows the comparison for the various systems with $N=15$ electrons. The TCF wave functions contain five composite fermions in each of the partitions. The top panel of the figure shows the spectrum for the flux corresponding to the incompressible state. The TCF incompressible state is the exact ground state of the four-body interaction. The energy of the neutral TCF mode, which contain an exciton in one of the three partitions, is also shown; it is separated from the ground state by a gap. A corresponding neutral mode can be identified also in the exact spectrum, indicated by blue colored dots. The quantum numbers predicted by the TCF model match the quantum numbers of the neutral mode in the exact spectrum. The overlaps improve with increasing angular momenta, i.e., with increasing distance between the quasiparticle and the quasihole forming the the exciton (the larger the angular momenta of TCF state, the larger is the distance between quasiparticle and quasihole).
The middle panel of figure \ref{Trial-4B-3N} shows the spectrum with one extra flux. The TCF states in this case have three quasiholes, one in each partition. These TCF states are the exact zero energy states of the Hamiltonian.
Finally, the bottom panel shows the spectrum when a flux is removed from the incompressible state, which produces a state with one quasiparticle in each partition. There appears to be a low energy mode whose quantum numbers closely match with the quantum numbers of the TCF states but the mode is not as well defined as the quasihole or neutral excitation mode. The overlaps decrease with increasing angular momenta. Since the excitations in the TCF states are all negatively charged, the average distance between excitations decreases with increasing angular momentum, which again is consistent with the observation that the agreement improves with increasing inter-quasiparticle distance. For small inter-particle separations (large angular momenta), the TCF theory also fails to predict the $L=6.5$ state which appears to be in the low energy mode in the exact spectrum. 

Figure \ref{Trial-4B-3Npm} shows the spectra for a system with $N=14$, described by TCF wave functions with unequal number of electrons in the three partitions. Given the finite system sizes, we have chosen specific systems which have a small number of quasiparticles and quasiholes. Removal of one flux and an electron from $(N,2Q)=(15,22)$ results in a state where there are two quasiparticles in one of the partitions. The exact and TCF spectra of such a system is shown in figure \ref{Trial-4B-3Npm}. Since all excitations are quasiholes, these states form exact zero energy states of the four-body interaction. Since both the quasiholes are in the same partition, the angular momentum counting of the low energy states is identical to that of the counting of the Laughlin state of $\nicefrac{N}{3}$ particles under $V_1$ interaction.
The middle panel of Figure \ref{Trial-4B-3Npm} shows the spectrum for a state with two quasiparticles and one quasihole distributed in three partitions; this is obtained from $(N,2Q)=(15,22)$ by removing 
 two fluxes and one electron. Since the system is composed of multiple excitations of different charges, there is no simple relation between angular momenta and inter-excitation distance. Even though there is no clear separation between low energy states and the bulk of the spectrum, there appears to be a very good agreement between the TCF trial wave functions and the lower energy states. Finally, the bottom panel of Figure \ref{Trial-4B-3Npm} compares the spectrum of a system with two quasiparticles in the same partition, obtained by adding one flux and one electron to the incompressible state. 

\begin{figure}
\includegraphics[scale=.6]{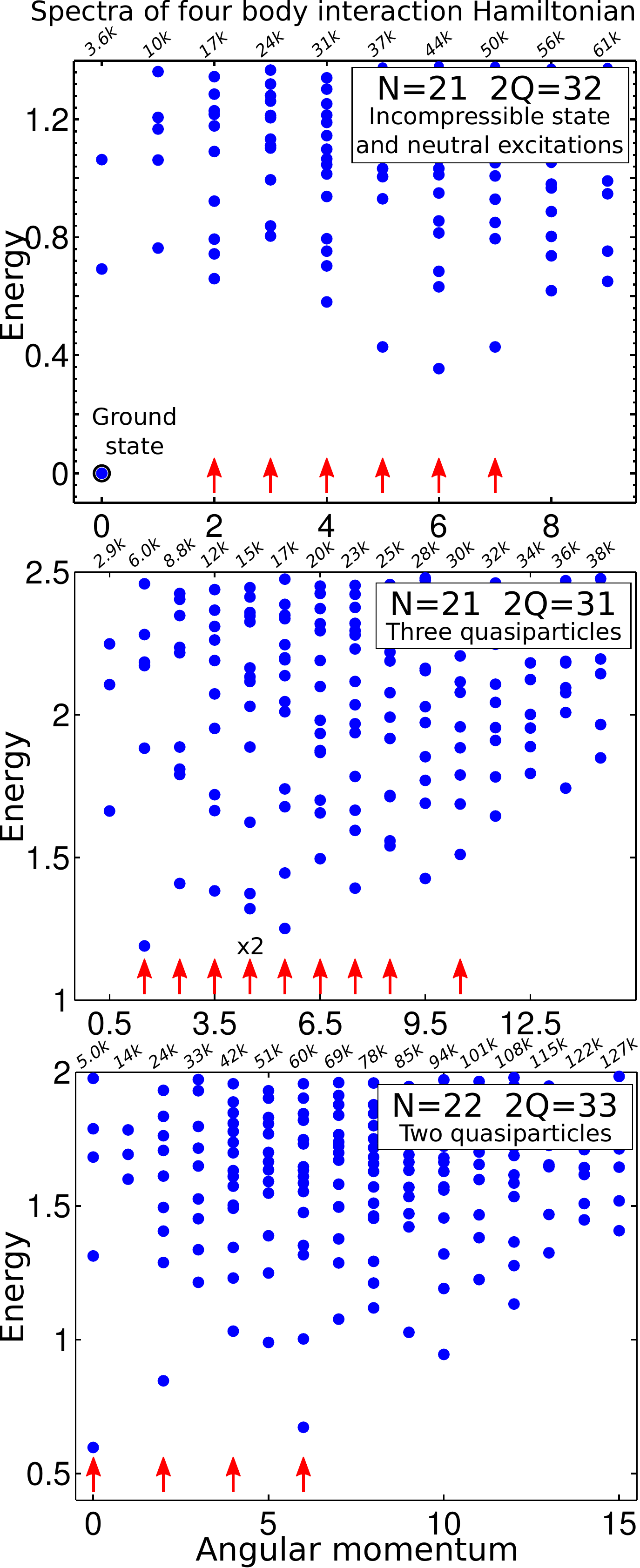}
\caption{Energy spectra of the four-body model Hamiltonian for certain systems with $N=21$ and $N=22$. The arrows indicate quantum numbers predicted by the TCF theory for the lowest band states ($\times 2$ denotes two states). The dimension of each angular momentum sector is shown at the top of each graph. The top panel corresponds to the system with an incompressible ground state; the middle and bottom panels contain three and two quasiparticles. In all the cases it is found that the predicted quantum numbers match the spectra in the limit where the quasiparticles are farthest separated.}
\label{4B-Larger}
\end{figure}

While it is computationally difficult to calculate overlaps between TCF states and exact states for larger systems, predictions for the quantum numbers of the low energy states can be compared with the exact spectra for larger systems. Figure \ref{4B-Larger} shows the spectrum for systems at and close to the incompressible state with $N=21$ particles. Arrows indicate the quantum numbers of the states predicted by the TCF theory. The spectra seem to match the predictions for large inter-particle separations, i.e. for large $L$ for the neutral mode (top panel) and for small $L$ for the quasiparticles. 

The picture emerging from these comparisons suggests that the TCF description is reasonably accurate for the four-body interaction ${\cal H}_4$. More specifically, it provides a correct counting of the number of states in the low energy band for situations where the quasiparticles and quasiholes are far separated, and the TCF wave functions are also an excellent approximation of the exact eigenstates under the same conditions. We believe that the deviations between the exact results and the TCF theory are due to finite system sizes; unfortunately, it is not possible to go to much larger systems than those studied here to test this assertion more conclusively.

\begin{figure}
\includegraphics[scale=.65]{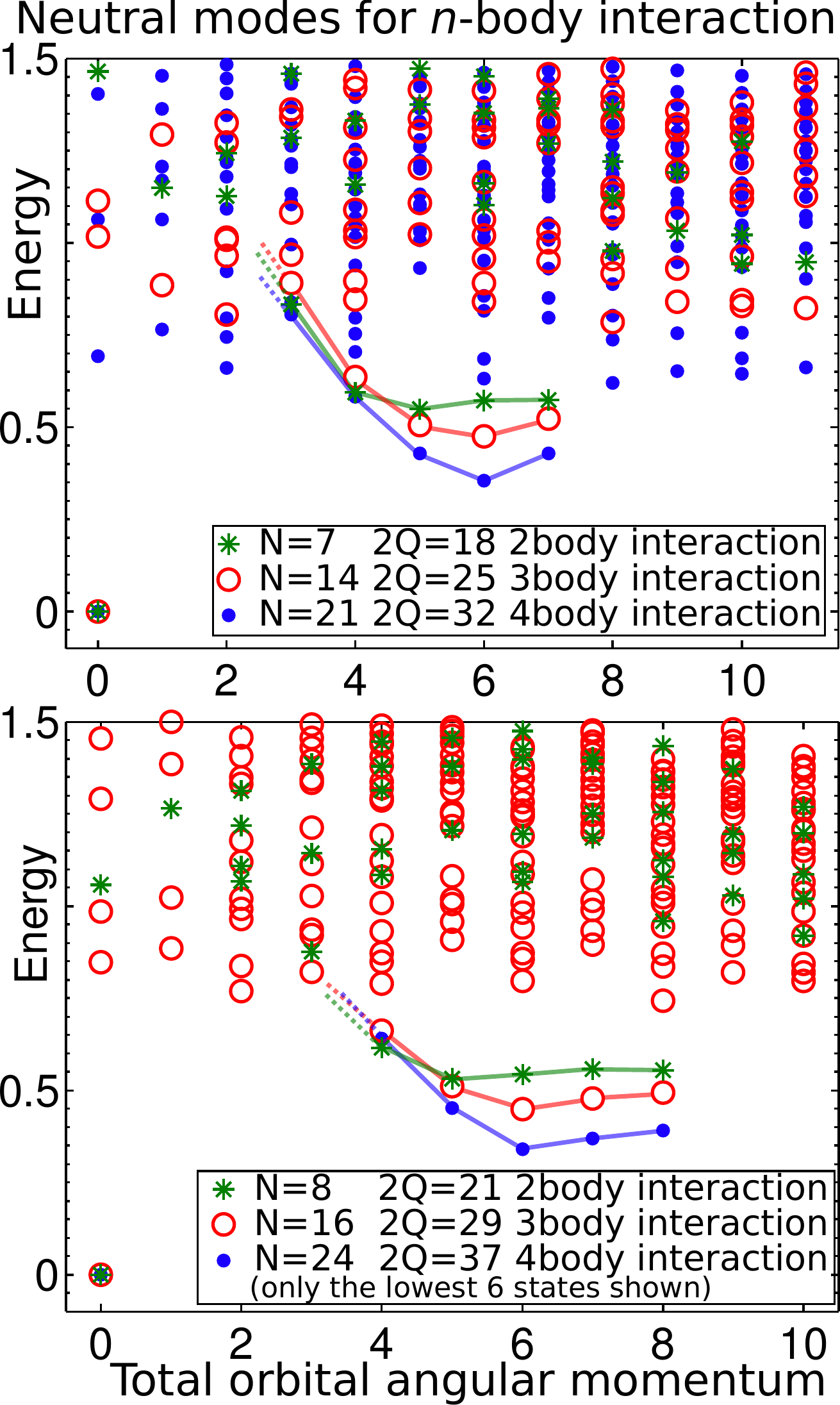}
\caption{Energy spectra of $n$-body model Hamiltonians $\mathcal{H}_2,\mathcal{H}_3,\mathcal{H}_4$ for which the Laughlin, Pfaffian and RR states are the highest density exact ground states. Each graph shows the energy spectra of $N$, $2N$ and $3N$ particles at fluxes $\frac{3}{1}N-3$, $\frac{2}{1}2N-3$ and $\frac{5}{3}3N-3$ respectively where incompressible states occur. The top panel has $N=7$ and the bottom panel has $N=8$. Similarity in the structures of the neutral modes provides a strong support to the notion that they correspond to multipartite states with one, two and three partitions respectively.}
\label{2-3-4bodyspectra}
\end{figure}

We close this section by making an interesting observation. The tripartite wave function suggests that the neutral mode of $3N$ particles with four-body interaction is analogous to the neutral mode of $N$ particles with 2-body interaction, and, in fact, also to the neutral mode of $2N$ particles for three-body interactions. The similarity can be explicitly seen in the spectra of the three interactions shown in figure \ref{2-3-4bodyspectra}. This is a direct evidence for the multipartite nature of the wave function. 

\section{Comparison with exact Coulomb solution: ground state}
\label{sec:comparewithCoulombGS}

Having shown that the TCF wave functions are reasonably good for the four-body interaction, we now test their validity for the Coulomb interaction. This section and the next are devoted to that issue. It is a priori far from obvious, and actually counter-intuitive, that the solutions of the four-body interaction should have anything to do with the solutions of the Coulomb interaction. Nonetheless, one can explicitly test if a connection exists. The current section presents comparisons for the incompressible ground state, and the next for the excitations.  In the absence of LL mixing, as assumed here, the Coulomb interaction is particle-hole symmetric, and therefore a trial wave function at $\nicefrac{2}{5}$ also implies a trial wave function at $\nicefrac{3}{5}$ (and vice versa), and one can choose to study either fraction. We remind the reader that `Coulomb interaction' refers to the second LL Coulomb interaction in this article.

\label{subsec:RR-BS-CF}

Several candidate wave function can be considered for the FQHE observed at filling fraction $2+\nicefrac{2}{5}$. 
The particle hole conjugate of the RR wave function occurs at a shift $-2$; it is constructed by producing the RR $\nicefrac{3}{5}$ wave function by exact diagonalization of the four-body interaction Hamiltonian, followed by particle hole conjugation. 
The Bonderson-Slingerland (BS) wave function  \cite{BondersonSlingerland2008,BondersonFeguinMollerSlingerland2009} occurs at shift $2$ and can be written as 
\begin{eqnarray}
\Psi_{\rm{BS}} &=& \mathcal{P}_{\rm{LLL}} \left\{ \Phi^*_2 \; \Phi^3_1 \; \rm{Pf}\left[ \frac{1}{z_\alpha - z_\beta}\right] \right\} \nonumber \\
&\approx& \rm{Pf}\left[ \frac{1}{z_\alpha - z_\beta}\right]\Psi^{\rm CF-bosons}_{\frac{2}{5}},
\end{eqnarray}
where $\Phi_n$ is the wave function with $n$ filled Landau levels and $\Psi^{\rm CF-bosons}_{\frac{2}{5}}\approx {\cal P}_{\rm LLL} \Phi_2^*\Phi_1^3$ is the Jain CF wave function for bosons at $\nu=2/5$. The Jain 2/5 wave function at shift $4$ is given by
\begin{equation}
	\Psi_{\rm{Jain}} = \mathcal{P}_{\rm{LLL}} \Phi_1^2 \Phi_2,
\end{equation}
and describes noninteracting composite fermions.
Finally, a BCF wave function at  $\nicefrac{3}{5}$ can be constructed as \cite{Sreejith11} 
\begin{equation}
\Psi_{\rm{BCF}} = \mathcal{A} \left [ \Psi_{\frac{3}{7}}(w_1\ldots , w_\frac{N}{2}) \Psi_{\frac{3}{7}}(z_1\ldots , z_\frac{N}{2}) \prod_{i,j=1}^{N/2} (z_i-w_j) \right ],
\end{equation}
where $z_i$, $w_i$ is an arbitrary partition of the particles into two equal parts and $\Psi_{\frac{3}{7}}$ is the Jain CF wave function at $\nicefrac{3}{7}$. Particle hole conjugate of this BCF function occurs at a shift of $-5$.

It is possible to rule out a candidate state as a possible explanation of the $2+\nicefrac{2}{5}$ FQHE by computing the exact Coulomb ground state at the corresponding shift. A necessary condition for the applicability of the trial wave function is that the exact Coulomb ground state at that shift have $L=0$ in the thermodynamic limit. A state with $L\neq 0$ represents quasiparticles or quasiholes of an incompressible FQHE state with a different shift. We note that the condition $L=0$ is not sufficient, however, because there can be more than one possibility at a given shift; a definitive confirmation of a theory requires that excitations also be explained by the theory. An instructive example in this context is the $\nicefrac{2}{5}$ FQHE in the lowest LL, where two proposals, namely the Gaffnian \cite{SimonRezayiCooper2007,Gaffnian} and Jain CF wave functions, have the same shift. In this case, the mere fact that the Coulomb ground state has $L=0$ is insufficient to discriminate between the two. However, the two models predict very different structure for quasiholes and quasiparticles, and a study of the excitations rules out the Gaffnian model  \cite{TokeJain09,Regnault09}.

\begin{figure}[h!]
\includegraphics[scale=.95]{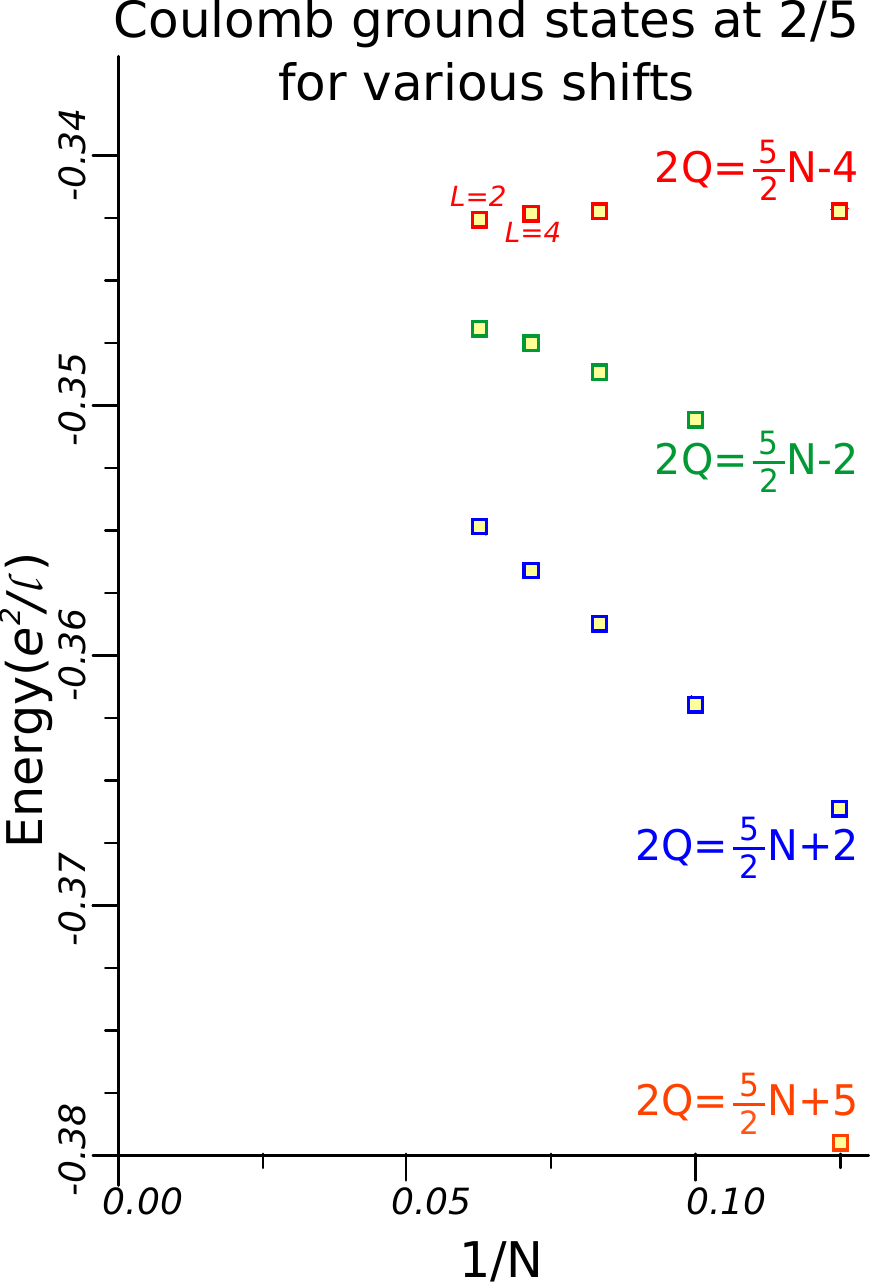}
\caption{ Ground state energies and angular momenta of $N$ electron interacting with the second Landau level Coulomb interaction on the sphere at various fluxes plotted against $\nicefrac{1}{N}$ for $N$ up to 16. The shifts $-2$, $2$ and $4$ and $-5$ correspond to Read-Rezayi, Bonderson-Slingerland, Jain, and the BCF states, respectively. The angular momentum of the ground state is shown whenever it is nonzero. The ground states at shift $4$ has a nonzero angular momenta, indicating that it represents excitations of some other incompressible state. The energies per particle are given in units of $\nicefrac{e^2}{\ell}$, where $\ell=\sqrt{\hbar c/eB}$ is the magnetic length, and include the interaction with the uniform positively charged background.}
\label{comparison-of-states}
\end{figure}

\begin{figure}
\includegraphics[scale=.5]{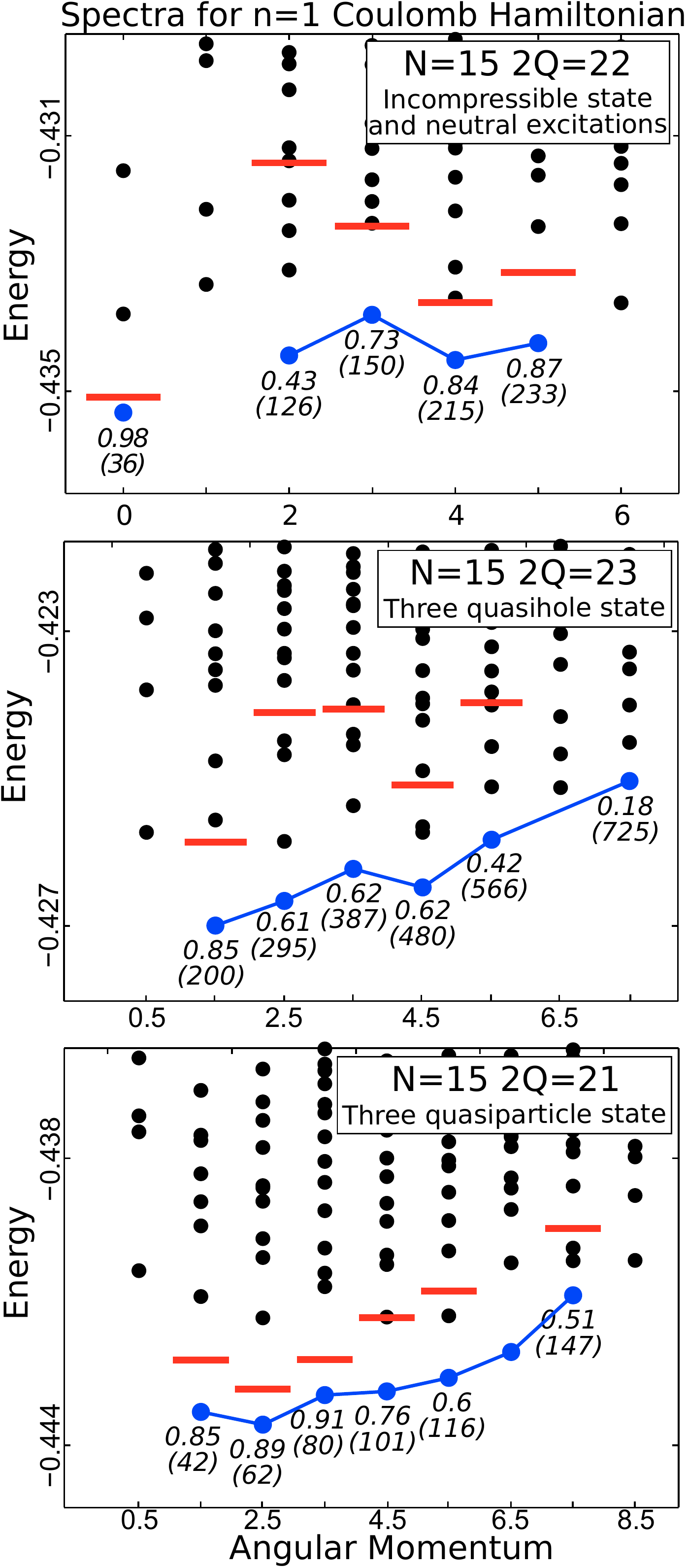}
\caption{Comparison of the TCF trial wave functions with the exact spectra of the second LL Coulomb interaction for systems with $N=15$ particles. The notations are similar to that of figure \ref{Trial-4B-3N} and \ref{Trial-4B-3Npm}. (top): An $L=0$ incompressible state and a clear neutral mode is formed at $2Q=22$. (middle): Overlaps between the low energy modes and the TCF states appear to decrease with increasing angular momenta (equivalently closer quasiholes). There is a gap between low energy modes and the bulk in the case of small angular momenta (farthest quasiholes) which vanishes when the quasiholes are close together. (bottom): A three quasiparticle mode in this scenario is similar to the quasiparticle mode in the four-body interaction. TCF states capture all except the state at $L=6.5$.}
\label{3N-Coulomb}
\end{figure}

\begin{figure}
\includegraphics[scale=.5]{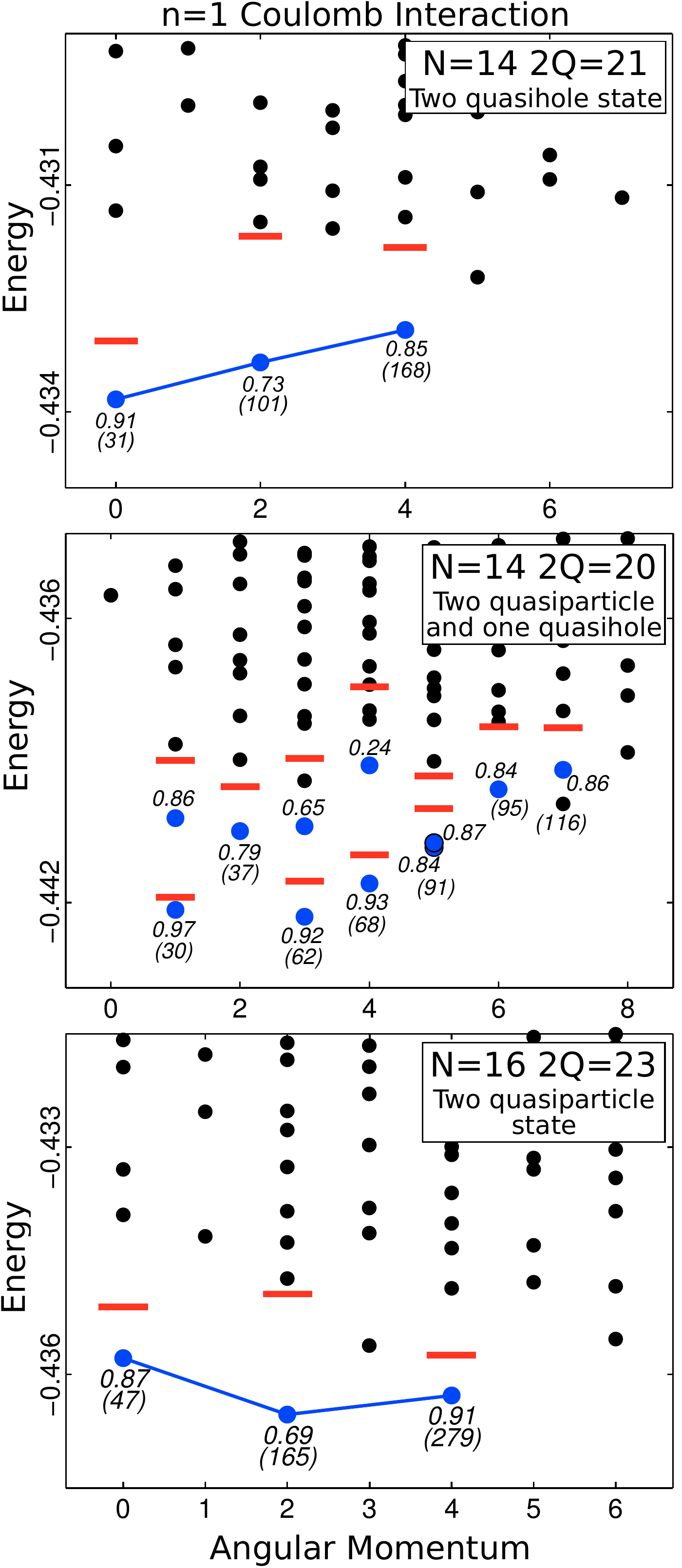}
\caption{Comparison of the exact spectrum and the TCF states for the cases where the number of particles is not a multiple of three. The meaning of various symbols is the same as in figures \ref{Trial-4B-3N}, \ref{Trial-4B-3Npm} and \ref{3N-Coulomb}. The states considered here are obtained from the incompressible state with $(N,2Q)=(15,22)$ shown in Fig.~\ref{3N-Coulomb} by 
removal of one electron and one flux (top);  by removal of one electron and two fluxes (middle); and addition of one electron and one flux (bottom).}
\label{3Npm-Coulomb}
\end{figure}

Figure \ref{comparison-of-states} shows the evolution of the actual Coulomb ground state as a function of the particle number $N$ for the shifts corresponding to the RR, BS, BCF and Jain wave functions. (Only one system is accessible to exact diagonalization at the BCF shift.) The energies per particle include the contributions from background-background and background-electron interactions. To minimize the shift dependence, 
the energies are rescaled by a factor of $\sqrt{\nu 2Q/N}$. We stress that this figure is not to be treated as a comparison between the energies; it is guaranteed that the exact Coulomb energy per particle will extrapolate to the same value in the thermodynamic limit independent of the shift, because the exact states at these shifts will only have order-one energy differences. However, for sufficiently large $N$, only one shift will produce a uniform $L=0$ ground state, whereas the states at nearby shifts will contain quasiparticles or quasiholes of this state and will in general have $L\neq 0$.  The exact Coulomb states at the RR, BS, and BCF shifts continue to have zero angular momenta for all systems that we have studied, indicating that the system sizes accessible to our study are not able to discriminate between them; they all remain viable candidates for the $2+\nicefrac{2}{5}$ FQHE. 

It is noteworthy that as the system size is increased, the angular momentum of the ground state at the shift $4$ changes to nonzero values, thereby ruling out the weakly-interacting CF-IQHE description for the $2+\nicefrac{2}{5}$ state. Same is true even for cases where finite thickness corrections are included\cite{ArekPRB80_2009}. This indicates that the physics of the second LL $\nicefrac{2}{5}$ FQHE is distinct from the lowest LL $\nicefrac{2}{5}$ FQHE.

\begin{table*}[h!]
\begin{tabular}{|c|c|c||c|c|c||c|c|c|}
\hline 
$N$ & $2Q$ & dim of L=0 & overlap & $E^{\rm Coulomb}_{\rm exact}$ & $E^{\rm Coulomb}_{\rm RR}$  & overlap & $E^{\rm Coulomb}_{\rm exact}$  & $E^{\rm Coulomb}_{\rm RR}$ \tabularnewline
 &  & subspace & ($w=0$) & ($w=0$) & ($w=0$) & ($w=3$) & ($w=3$) & ($w=3$)\tabularnewline
\hline 
15 & 22 & 36 & 0.9836 & -0.6490 & -0.6486 & 0.9801 & -0.4607 & -0.4604\tabularnewline
\hline 
18 & 27 & 319 & 0.9369 & -0.6480 & -0.6471 & 0.8995 & -0.4625 & -0.4618\tabularnewline
\hline 
21 & 32 & 3603 & 0.8990 & -0.6469 & -0.6457 & 0.9316 & -0.4631 & -0.4625\tabularnewline
\hline 
24 & 37 & 50866 & 0.8100 & -0.6463 & -0.6449 & 0.8792 & -0.4639 & -0.4631\tabularnewline
\hline 
\end{tabular}
\caption{Comparison between RR state and the second Landau level Coulomb ground state on the sphere for two different quantum well widths $w=0$ and $w=3$ (quoted in units of the magnetic length). $N$ is the number of electrons and $2Q$ is the number of flux quanta penetrating the surface of the sphere; the full dimension of the $L=0$ subspace is also given. For each case, the table gives the Coulomb energies of the exact and the RR states, $E^{\rm Coulomb}_{\rm exact}$ and $E^{\rm Coulomb}_{\rm RR}$, respectively, as well as the overlaps between them. 
Energies per particle are given in units of $\nicefrac{e^2}{\epsilon\ell}$, where $\ell=\sqrt{\hbar c/eB}$ is the magnetic length and $\epsilon$ is the dielectric constant of the background, and include the interaction with the uniform positively charged background. Finite width calculations use the model described in Ref.~[\onlinecite{WojsQuinn2007}]. The overlaps for the 15 and 18 particle systems at $w=0$ were previously given in Ref.~[\onlinecite{ReadRezayi1999}].}
\label{table1}

\end{table*}

We next proceed to compute the overlap between the RR state and the exact Coulomb state and also compare their Coulomb energies as a function of $N$, shown in Table \ref{table1}.  The overlaps are quite large and provide nontrivial support for the RR state. 

We have also investigated if the ground state evolves adiabatically (without gap closing) when the interaction is changed from the four-body model interaction to the second LL Coulomb. For this purpose, we diagonalize the Hamiltonian
\begin{equation}
	\mathcal{H}(\lambda) = (1-\lambda)\frac{\mathcal{H}_{\rm Coulomb} - E_{\rm Coulomb}}{\Delta_{\rm Coulomb}} + \lambda\frac{\mathcal{H}_4 - E_4}{\Delta_4}\label{eq:interpolationHamiltonian}
\end{equation}
as a function of the parameter $\lambda$. The parameters $E_4$ and $E_{\rm{Coulomb}}$ are the energies of the lowest energy state in the for the $\mathcal{H}_4$ and $\mathcal{H}_{\rm{Coulomb}}$ spectrum. Scaling factors $\Delta_{\rm{Coulomb}}$ and $\Delta_4$ are chosen to be of the order of the gap between the ground state and the first excited state in each interaction. If there is no clear gap in the spectrum, the $\Delta$s are chosen to be of the order of the gap between the lowest energy state and the first excited state in the same angular momentum sector.
This Hamiltonian gives the four-body interaction in $\lambda=0$ limit and the second LL Coulomb in the $\lambda=1$ limit (up to an overall shift and a scaling factor). The results, shown in the next section (see figure \ref{15AD}) along with the evolution of the excitations, indicate that the gap does not close, thus providing further support for the RR wave function.
We have not carried out similar calculations for the BS and the BCF wave functions, which are outside the scope of our present paper. 

\section{Comparison between TCF wave functions and exact Coulomb solutions: Excitations}
\label{sec:TCF-CoulombExcitations}

We next compare the TCF excitations with the actual Coulomb excitations.  Exactly as done previously for ${\cal H}_4$, we obtain the spectra and eigenstates by diagonalizing the second LL Coulomb interaction (i) in the full Hilbert space and (ii) in the TCF basis, and then compare the two results.  Figure \ref{3N-Coulomb} shows the comparison for 15 particles for the ground state, neutral excitations, quasiholes and quasiparticles; this figure is analogous to the previous Fig. \ref{Trial-4B-3N}. The incompressible TCF state (RR state) has high overlap with the exact Coulomb ground state and predicts the quantum numbers of the neutral mode correctly. However the neutral mode of the Coulomb system is not as clearly formed as it is for the four-body interaction. As for the four-body interaction, we find that the overlaps in general are better when the quasiparticle and the quasihole of the neutral exciton are far separated.  The remaining panels of Figure \ref{3N-Coulomb} test the validity of the TCF model for quasihole and quasiparticle excitations (center and bottom panels, respectively). The agreement between the TCF states and the exact spectra is poor when the quasiholes or quasiparticles are close together, but improves when they are far separated (i.e. at small angular momenta). 
Figure~\ref{3Npm-Coulomb} compares the trial wave functions with the exact spectra for cases where the number of particles is not a multiple of $3$. This figure shows that the TCF model is reasonable for the Coulomb solution, though not as accurate as it is for the four-body interaction. 

\begin{figure*}
\includegraphics[scale=.45]{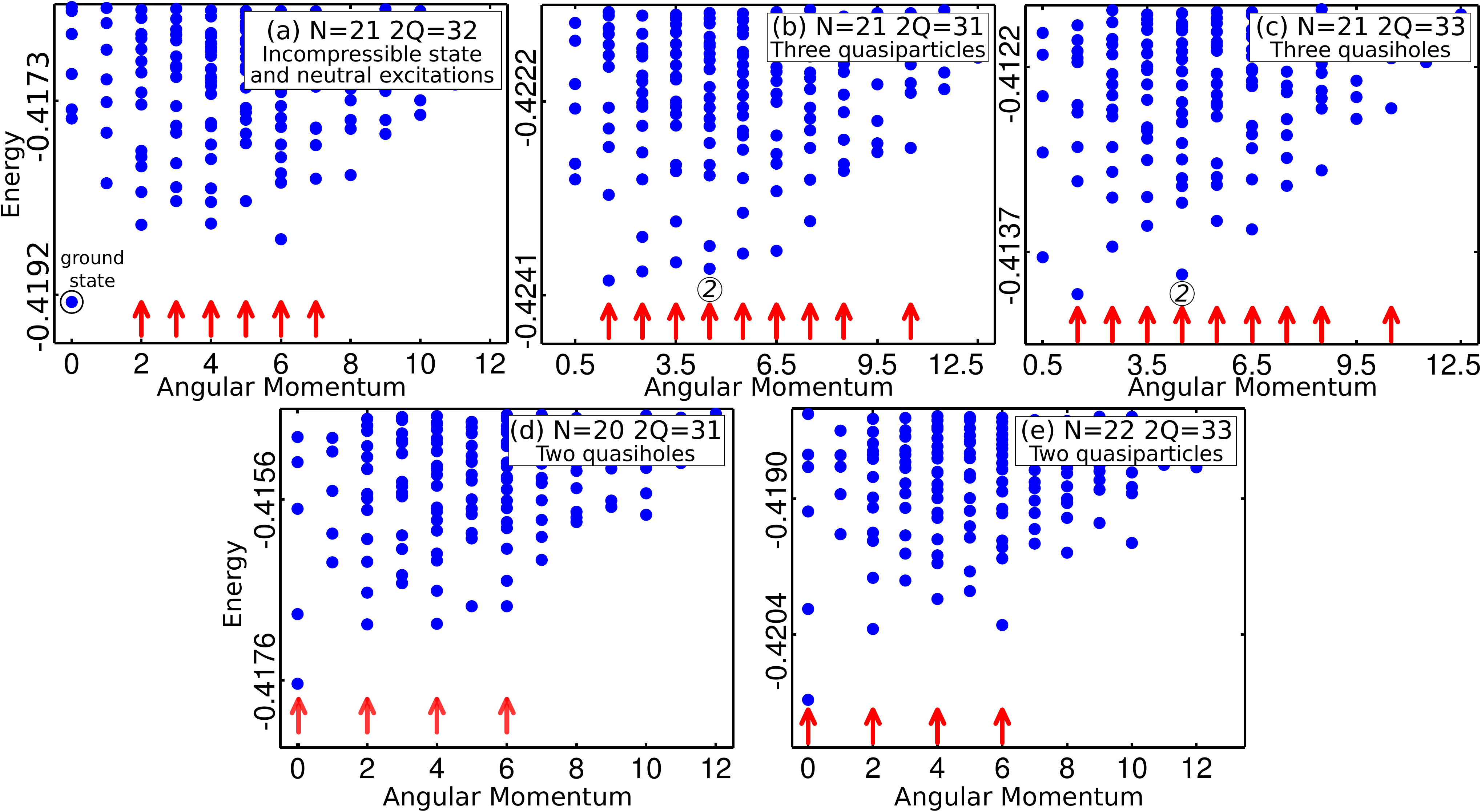}
\caption{Second LL Coulomb spectra of systems at and in the vicinity of the incompressible state with $(N,2Q)=(21,32)$. The arrows indicate the angular momentum quantum numbers predicted by the TCF model, where the encircled 2 indicates a doublet.}
\label{Coulomb-Larger}
\end{figure*}

Figure \ref{Coulomb-Larger} shows the spectra of incompressible state and excitations of a system of $N\sim 21$ particles, together with predictions for quantum numbers from the TCF model. While the TCF captures several features of the spectra correctly, in general there is a poor agreement with the exact spectrum. The neutral mode is not clearly formed in panel (a), and the number of predicted states do not appear to form a low-energy band. The TCF model does capture some features, however. 
Counting of the low energy states at $2Q=31$ (figure \ref{Coulomb-Larger} (b)) is correctly predicted in the small angular momentum sectors (large inter quasiparticle distance). The absence of low energy states in the odd angular momentum sectors in figure \ref{Coulomb-Larger} (d and e) is also consistent with the TCF model.

\begin{figure}
\includegraphics[scale=.5]{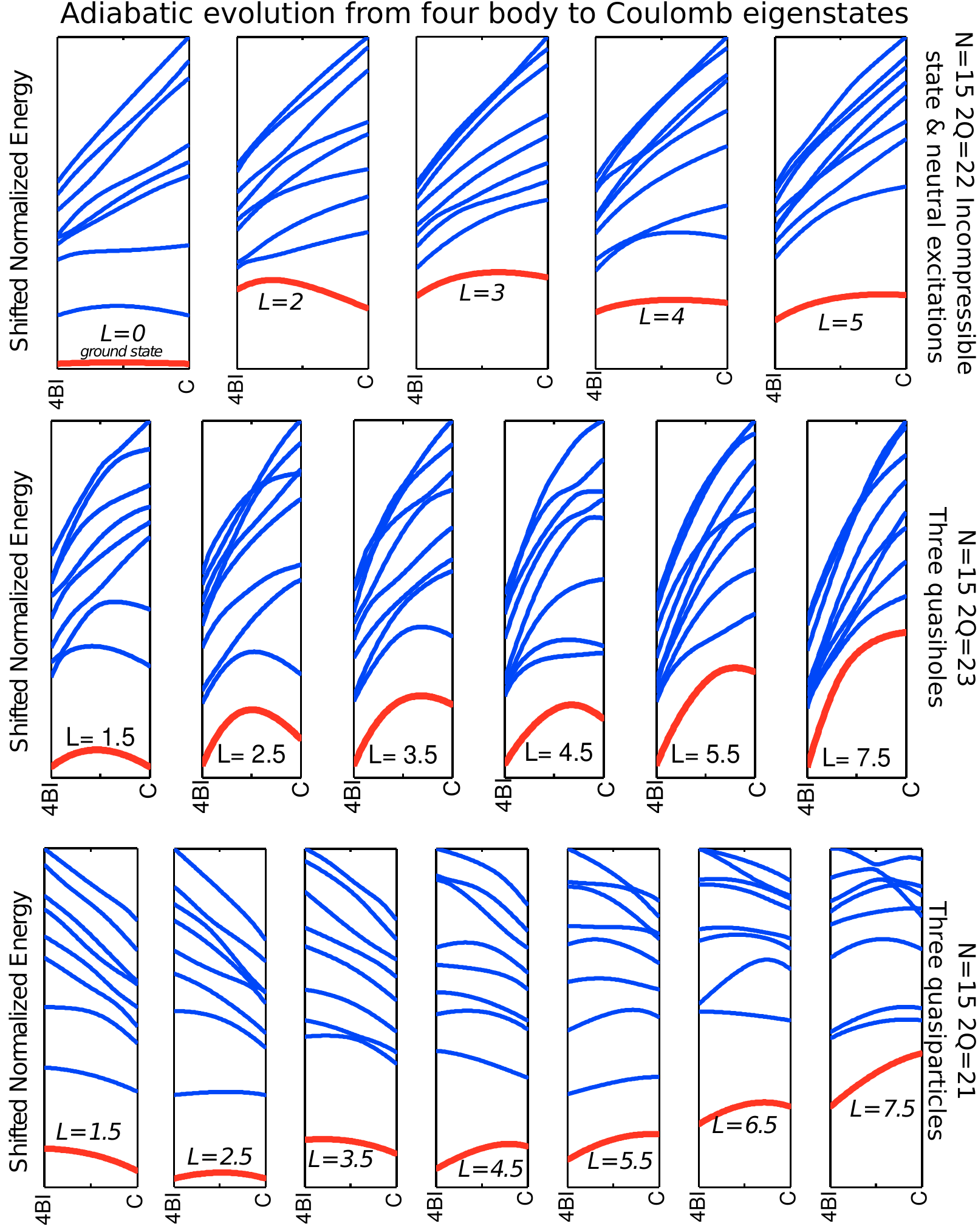}
\caption{Evolution of states as the Hamiltonian is tuned from four-body model Hamiltonian to the Coulomb Hamiltonian for $N=15$. Each panel shows the low energy states in a given $L$ sector. The top panels show evolution for the ground state and neutral excitons; the middle panels for three quasihole states, and the bottom panels for three quasiparticle states. 
The red line shows the evolution of the low energy states and blue lines show the higher energy states; the absence of any crossing between the red and the blue lines indicates that the Coulomb solutions are adiabatically connected to the solutions of the four-body Hamiltonian. Note that $L=6.5$ state in the three quasiparticle system, which appears to be a part of the low energy mode, is not predicted by the TCF theory.}
\label{15AD}
\end{figure}

\begin{figure}
\includegraphics[scale=.5]{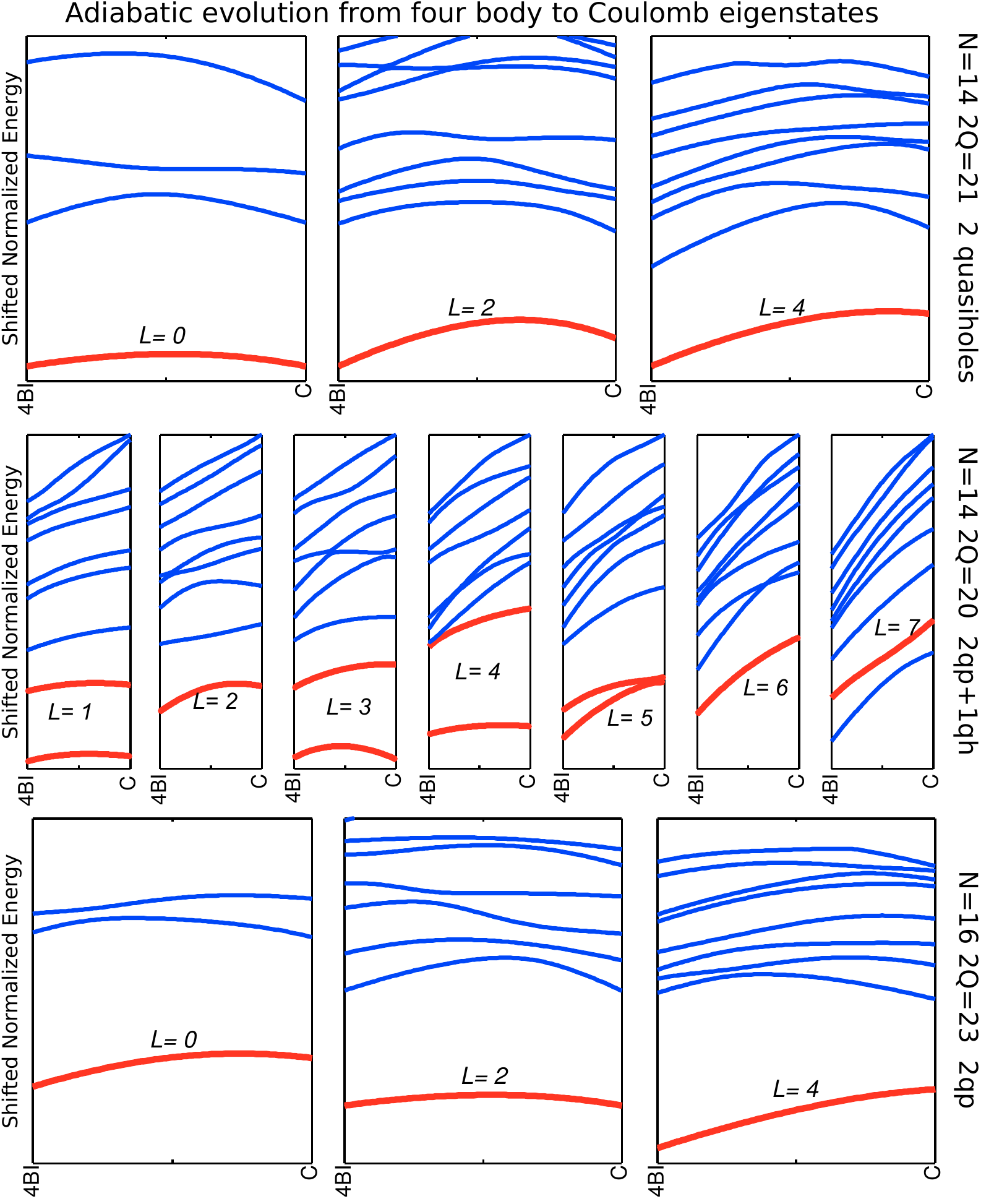}
\caption{Evolution of states as the Hamiltonian is tuned from four-body model Hamiltonian to the Coulomb Hamiltonian for the cases where $N$ is not a multiple of $3$. There is a clear adiabatic connection between low energy states of the two quasihole states (top panels) and two quasiparticle states (bottom panels).  In the center panels, all the angular momenta except $L=4$ show adiabatic connection. Note that in the middle panel, for $L=7$, we have shown the first excited state in red because this state has a higher overlap with the TCF wave function.}
\label{14-16AD}
\end{figure}

Finally, Figs. \ref{15AD} and \ref{14-16AD} show the evolution of the ground state as well as various excitations for the model interaction in Eq. \ref{eq:interpolationHamiltonian} which interpolates between the four-body interaction and the Coulomb interaction. The relevant eigenstates of the four-body interaction adiabatically evolve into the low energy Coulomb eigenstates without any gap closing within that angular momentum sector. We believe that these adiabatic evolutions make a strong case for a connection between the four-body and the second LL Coulomb Hamiltonians. 

\section{Conclusion}
\label{sec:Conclusion}

We have investigated in this paper the tripartite wave functions for the second LL filling of $2+3/5$. These reduce to the RR wave function for the ground state and quasihole excitations, but also provide a model for the neutral and quasiparticle excitations. The excitations are modeled through the standard CF excitations in the individual partitions. 

We have studied the plausibility of the TCF wave functions for the second LL Coulomb interaction as well as a four-body interaction. Of course, only the Coulomb interaction is relevant to experiment. The reason for studying the four-body interaction is that it provides another route to testing the validity of the TCF wave function for the Coulomb interaction in a two step process: by showing that the TCF wave functions are accurate for the four-body interaction, and then establishing adiabatic continuity to the Coulomb interaction. This method has proved useful in the studies of the MR state and also for the topological exciton that occurs in a paired CF state with an odd number of composite fermions \cite{MorfDasSarma2010,Sreejith11}. 

For the four-body interaction the TCF model is, by construction, exact for the ground state and quasiholes. We find that it is quite satisfactory for the neutral excitations, and also for collections of quasiparticles and quasiholes when they are well separated. 

For the Coulomb interaction, the RR ground state is quite accurate, with fairly high overlap with the exact Coulomb state even for 24 electrons. The situation for the quasiparticles, quasiholes and neutral excitations is less convincing, however. 
For $N=15$ electrons, the overlaps and counting of the TCF excited states closely match that found in the exact Coulomb spectra, and there is also an adiabatic continuity between the four-body spectrum and the Coulomb spectrum for the low energy states. However, as the system size is increased to $N=21$, the neutral excitations, quasiparticles and quasiholes seem to merge into the continuum of the spectrum, making it impossible to identify these modes and compare the counting of the states.  Overall, while these results lend general support to the RR / TCF physics at $2+3/5$, further work will be necessary for an unambiguous confirmation. The Bonderson-Slingerland and the bipartite CF states also remain viable candidates. A further study of their excitations will be necessary decisively to distinguish between these proposals. 

We finally note that experimental measurements of local quasiparticle charge and the presence of upstream neutral modes can also help distinguish between the various proposals. The BCF proposal produces quasiparticles with charge $e/10$, as opposed to a charge of $e/5$ predicted by RR and BS constructions\cite{BondersonFeguinMollerSlingerland2009}. The BCF and RR predict, for an ideal unreconstructed edge, upstream neutral edge modes \cite{Wen-edge} at $2+2/5$ but none at $2+3/5$ (because $2+2/5$ is obtained by particle hole conjugation of $2+3/5$), whereas BS implies upstream neutral modes at both $2+2/5$ and $2+3/5$ (because this wave function involves reverse flux attachment \cite{Wu93}).

\section*{Acknowledgements}

We acknowledge financial support from the NSF under grant no. DMR-1005536 (JKJ), the DOE under Grant No. DE-SC0005042 (GJS and YHW), and the Polish NCN grant 2011/01/B/ST3/04504 and EU Marie Curie Grant PCIG09-GA-2011-294186 (AW). YHW thanks N. Regnault for help on calculating Jack polynomials. GJS thanks E.~Ardonne and T.~H.~Hansson for discussion on general quasihole wavefunctions of RR and BS states. We thank Research Computing and Cyberinfrastructure, a unit of Information Technology Services at Pennsylvania State University, as well as Wroclaw Centre for Networking and Supercomputing and Academic Computer Centre CYFRONET, both parts of PL-Grid Infrastructure for providing high-performance computing resources and services used for the computations in this work. 

\begin{appendix}
\section{Multipartite composite fermion functions}
The BCF and TCF wave functions can be straightforwardly generalized to the case of multipartite CF wave functions. A general incompressible wave function containing $m$ partitions, each with $n$ filled $\Lambda$Ls, has the form 
\begin{align}
\Psi_{m,n,p}(z_1,..,z_{Nm})&=\mathcal{A}  \left[ \prod_{j=0}^{m-1} \psi_\frac{n}{2pn+1} \left(\{z_{jN+i}\}_{i=1,..,N}\right) \right. \nonumber\\
&\left. \prod_{k<l=0}^{m-1} \prod_{a,b=1}^N \left(z_{kN+a}-z_{lN+b}\right) \right]
\end{align}
In the above wave function, coordinates in the different CF partitions are correlated by cross terms of single power similar to the TCF and BCF states.
The above multipartite wave functions represent filling fractions
\begin{equation}
	\nu=\frac{nm}{1+(2p+m-1)n} \label{MCFfillingfraction}
\end{equation}
and in the spherical geometry, incompressible states of this function occur at flux values
\begin{equation}
2Q=\frac{N}{\nu}-(n+2p)
\end{equation}
The local charge of excitations can be calculated in a manner similar to that of the TCF states. A single flux through the state has a total charge of $\nu e$, but leads to formation of $n$ quasiholes in each of the $m$ partitions. Therefore a single localized excitation has a charge of
\begin{equation}
\frac{e}{1+(2p+m-1)n},
\end{equation}
The bipartite Pfaffian wave function corresponds to the parameters $(m,n,p)=(2,1,1)$ and the $k=3$ RR wave function corresponds to $(3,1,1)$. Other tripartite CF functions of the form $(3,n,p)$ occur at filling fractions $2/3,9/13,3/7,1/3$ etc. Bipartite states of class $(2,2,1)$ and $(2,3,1)$ which correspond to filling fraction $\nu=4/7$ and $\nu=3/5$ were studied in Ref [\onlinecite{Sreejith11}].
\end{appendix}

\end{document}